\documentclass[aps,prd,superscriptaddress,nofootinbib,eqsecnum,twocolumn]{revtex4-1}

\pdfoutput=1

\usepackage{amsfonts}
\usepackage{amsmath}
\usepackage{amssymb}
\usepackage{graphicx,color}
\usepackage{xcolor}
\usepackage{float}
\usepackage{hyperref}
\usepackage{subfigure}
\usepackage{mathtools,slashed}
\usepackage{blindtext}
\usepackage{longtable}
\usepackage{dcolumn}
\usepackage{bm}
\usepackage{appendix}
\usepackage{multirow}
\usepackage{color}
\usepackage{soul}
\usepackage{lipsum}

\begin{document}

\title{Constraining the Barbero-Immirzi parameter from the duration
  of inflation in loop quantum cosmology}

\author{L. N. Barboza} 
\email{lauziene.barboza@gmail.com}
\affiliation{Instituto de F\'{\i}sica, Universidade Federal Fluminense, 
Avenida General Milton Tavares de Souza s/n, Gragoat\'a, 24210-346 Niter\'oi, 
Rio de Janeiro, Brazil}

\author{G. L. L. W. Levy}
\email{guslevy9@hotmail.com}
\affiliation{Departamento de F\'{\i}sica Te\'orica, Universidade do Estado do 
Rio de Janeiro, 20550-013 Rio de Janeiro, RJ, Brazil}

\author{L. L. Graef} 
\email{leilagraef@id.uff.br}
\affiliation{Instituto de F\'{\i}sica, Universidade Federal Fluminense, 
Avenida General Milton Tavares de Souza s/n, Gragoat\'a, 24210-346 Niter\'oi, 
Rio de Janeiro, Brazil}

\author{Rudnei O. Ramos} 
\email{rudnei@uerj.br}
\affiliation{Departamento de F\'{\i}sica Te\'orica, Universidade do Estado do 
Rio de Janeiro, 20550-013 Rio de Janeiro, RJ, Brazil}

\begin{abstract}

We revisit the predictions for the duration of the inflationary phase
after the bounce in loop quantum cosmology. We present our analysis
for different classes of inflationary potentials that include the
monomial power-law chaotic type of potentials, the Starobinsky and the
Higgs-like symmetry breaking potential with different values for  the
vacuum expectation value. Our setup can easily be extended to other
forms of primordial potentials than the ones we have
considered. Independently on the details of the contracting phase, if
the dynamics starts sufficiently in the far past, the kinetic energy
will come to dominate at the bounce, uniquely determining the
amplitude of the inflaton at this moment. This will be the initial
condition for the further evolution that will provide us with results
for the number of {\it e}-folds from the bounce to the beginning of
the accelerated inflationary regime and the subsequent duration of
inflation.  We also discuss under which conditions each model
considered could lead to observable signatures on the spectrum of the
cosmic microwave background or else be excluded for
not predicting a sufficient amount of accelerated expansion. A first
analysis is performed considering the standard value for the
Barbero-Immirzi parameter, $\gamma \simeq 0.2375$, which is obtained
from black hole entropy calculations. In a second analysis, we
consider the possibility of varying the value of this parameter, which
is motivated by the fact that the Barbero-Immirzi parameter can be
considered a free parameter of the underlying quantum theory in the
context of loop quantum gravity. {}From this analysis, we obtain a
lower limit for this parameter by requiring the minimum amount of
inflationary expansion that makes the model consistent with the 
cosmic microwave background observations. When constraining the Barbero-Immirzi parameter, 
we have made use of the results for the scalar of curvature perturbations 
derived in loop quantum cosmology assuming the dressed metric approach and considering 
the Bunch-Davies vacuum as the initial condition for the perturbations in the 
contracting phase. Such choice provides basically the same results as 
considering the fourth-order adiabatic vacuum state 
at the bounce as an initial condition in the context of the dressed metric approach.

\end{abstract}

\maketitle

\section{Introduction}
\label{intro}

Inflation is the most popular paradigm for early universe cosmology, 
but it is not the only one. Indeed, it has been extremely successful 
from a phenomenological point
of view. It predicted the spatial flatness of the Universe, the
homogeneity seen in the cosmic microwave background (CMB),
besides suggesting a causal explanation for the origin of its
anisotropies, providing the first theory for the origin of the
large-scale structure of the Universe based on fundamental
physics. Although requiring a certain amount of fine-tuning on its
constants~\cite{Linde:2007fr,Vazquez:2018qdg,Kofman:2002cj} the
inflationary scenario predicts correctly the primordial power spectra,
whose evolution determines the temperature fluctuations in the 
  CMB and the formation of the large-scale structure of the
Universe~\cite{Vazquez:2018qdg,Planck:2018jri}, having been developed
before most of the data we now have was on hand. 

However, despite the success of inflation, in the classical theory of
general relativity (GR), all scalar field models of inflation
experience the big bang singularity that is
inevitable~\cite{Borde:1993xh,Borde:2001nh}. The singularity problem
is the result of an extrapolation of GR beyond the limits where
the theory is well justified. This implies additional difficulties
in the definition of initial conditions due to the absence of a
regular surface where those initial conditions can be established.
Another related problem in the context of inflation in GR is the
fact that there is a very limited value for the number of inflationary
{\it e}-folds which makes the model theoretically consistent. We know
that in order to be compatible with observations, the number of {\it
  e}-folds during inflation should be at least $60$ or so.  Meanwhile,
in some cases, the predicted number of {\it e}-folds is more than
$70$~\cite{Martin:2013tda}.  In those cases, the scale of the
fluctuations which are today observed in the CMB was smaller
than the Planck length at the starting of inflation. As a result, the
usual semiclassical treatment during inflation is questionable, which
is known as the trans-Planckian
problem~\cite{Brandenberger:2009rs,Brandenberger:2012aj}.  These
issues with the consistency of inflation also motivate to consider
scenarios with a well-defined preinflationary dynamics, free of
singularities~\cite{Zhu:2017jew,Shahalam:2017wba,Shahalam:2021wyo}. In
these scenarios, cosmological perturbations  may reach the beginning
of inflation in a quantum state excited relative to the Bunch-Davies (BD) 
vacuum, affecting the power spectra, thus potentially leaving marks on
the CMB~\cite{Martin-Benito:2021szh}. In order to address these
important issues, one must consider a scenario of quantum gravity
acting in the high energy regime. The scenario we consider here lies
in the context of loop quantum gravity~(LQG).

LQG proposes a  quantization formalism based on a
nonperturbative and background-independent formulation of the
geometric degrees of
freedom~\cite{Ashtekar:2011ni,Barrau:2013ula,Agullo:2016tjh,Bojowald:2006da,Ashtekar:2003hd,Ashtekar:2006rx,Ashtekar:2009mm}. In
its canonical formulation, it is aimed to respect the general
covariance of Einstein’s theory.  In particular, in GR the
Hamiltonian is a linear combination of constraints which, via Poisson
brackets, generates diffeomorphism transformations, the fundamental
symmetry of the theory~\cite{ElizagaNavascues:2020uyf}.  LQG, on
the other hand,  adopts the quantization scheme proposed by Dirac for
systems with constraints, which consists in requiring that those
constraints are satisfied at the quantum level on the physical states
of the system. The geometric degrees of freedom  are described in 
  LQG by pairs of canonical variables, the Ashtekar variables, which
consist of the components of a densitized triad and a gauge
connection. The quantization is obtained through holonomies of the
connections and fluxes of the densitized triads.  

Loop quantum cosmology (LQC) is the symmetry reduced version of
  LQG~\cite{Ashtekar:2003hd}, applied to cosmological
models~\cite{Agullo:2016tjh,Bojowald:2006da,Banerjee:2011qu}. Taking
into account such quantum geometric effects described by LQG in
cosmological scenarios, Einstein's equations maintain an excellent
degree of approximation in the low curvature regime. However, in the
Planck regime they undergo major changes, driving a nonsingular
bounce due to repulsive quantum geometry
effects~\cite{Ashtekar:2011ni,Ashtekar:2009mm}, then naturally solving
the big bang singularity problem. Therefore,  in the LQC
framework, for matter satisfying the usual energy conditions, whenever
a curvature invariant grows near the Planck scale in LQC, the
effects of quantum geometry dilute it~\cite{Ashtekar:2011ni}.

In LQC models with a kinetic energy dominated bounce, as the
ones we are going to consider, an inflationary phase almost inevitably
follows the bounce phase (see, e.g.,
Refs.~\cite{Zhu:2017jew,Shahalam:2017wba,Li:2018fco,Sharma:2018vnv,Li:2019ipm,Shahalam:2019mpw}). 
This is true whenever there is an inflaton field, with an appropriate potential,  
coupled to the gravitational field (otherwise, if there is only radiation and/or 
cold dark matter in the bounce this does not happen). The
duration of this inflationary phase is quantified by the number of
{\it e}-folds~\cite{Barrau:2020nek,Bedic:2018gqu,Graef:2018ulg,Barboza:2020jux}.
As it is well known~\cite{Guth:1980zm}, the inflationary phase must
last at least around $60$ {\it e}-folds or so in order to solve the
problems of the standard cosmological model. However, in LQC, as
shown in Ref.~\cite{Agullo:2013ai}, the bounce and preinflationary
dynamics leaves  imprints on the spectrum of the CMB. In
Ref.~\cite{Zhu:2017jew} it was shown that,  in order to be consistent
with observations, the Universe in LQC must have expanded at
least around $141$ {\it e}-folds from the bounce until today. This is
so because LQC can lead to scale-dependent features in the
  CMB spectrum, and the fact that we do not observe them today means
that they must have been well diluted by the postbounce expansion of
the Universe. This implies an extra number of inflationary {\it
  e}-folds in LQC, given by $\delta N \sim
21$~\cite{Zhu:2017jew}. On the other hand, if the number of extra
inflationary {\it e}-folds is much higher than this value the features
imprinted in the CMB spectrum due to the LQC effects would
be overly diluted and, in this case, LQC cannot be directly
tested even by forthcoming experiments.

Such theoretical results motivate an investigation of the number of
{\it e}-folds in models of LQC. This can be performed
consistently with initial conditions defined either in the bounce, as
done, e.g., in Refs.~\cite{Ashtekar:2011rm,Zhu:2017jew}, or in a
contraction phase before the bounce, as considered in
Refs.~\cite{Linsefors:2013cd,Linsefors:2014tna,Bolliet:2017czc,Martineau:2017sti},
for example; in particular, with the authors of the latter references
claiming that taking initial conditions in the far past in the
contracting phase should be the appropriate approach to study the
probability of inflation after the bounce.  Both approaches for
choosing the initial conditions to investigate the probability of
inflation in LQC were applied to many different
potentials. {}For instance, power-law potentials were considered in
Refs.~\cite{Linsefors:2013cd,Linsefors:2014tna,Bolliet:2017czc,Martineau:2017sti,Shahalam:2017wba,Chen:2015yua,Barboza:2020jux},
alpha-attractor potentials in Ref.~\cite{Shahalam:2019mpw}, monodromy
potentials with a modulation term in Ref.~\cite{Sharma:2018vnv} and
also chaotic and Starobinsky potentials in the framework of standard
and modified LQC models~\cite{Li:2019ipm}. In particular, in the
work of Refs.~\cite{Graef:2018ulg,Barboza:2020jux}, in addition of
considering  different classes of potentials, the effect of radiation
as an additional ingredient of the energy density budget around the
bounce has been considered, showing that the duration of inflation is
dependent not only on the inflaton potential but also on the amount of
radiation in the Universe prior to the start of the inflationary
regime. 

In the present paper, we show that there are  well-defined values for
the inflaton amplitude at the bounce, which can be estimated
analytically and depending only on the form of the primordial inflaton
potential. Then, we provide explicit results, also analytically, for
the duration of the preinflationary phase and for the inflaton
amplitude at the beginning of the inflationary regime. This then
allows us to explicitly obtain the total number of {\it e}-folds that
inflation will have. Our results are sufficiently general such that it
can easily be extended to many different forms of the primordial
inflaton potential other than the ones we have focused in this paper. 

In the analysis we perform in the first part of this paper, we
consider the standard value for the {\it Barbero-Immirzi parameter},
$\gamma$, which is the one arising from the calculation of the entropy
of black holes~\cite{Barrau:2018rts}. Then in the second part, we
 consider the Barbero-Immirzi parameter as a free variable of the
theory, and we obtain its value  by demanding each model to predict 
the correct spectrum of CMB. The choice of varying this parameter 
is motivated by the fact that the Barbero-Immirzi parameter  is, indeed,
a coupling constant with a topological term in the action of gravity,
with no consequence on the classical equations of motion.\footnote{{}For 
further discussion in this topic, see, e.g.,
Refs.~\cite{Armas:2021yut,Asante:2020qpa,Asante:2021zzh,Perlov:2020cgx,Broda:2010dr,Mercuri:2009zt,Gates:2009pt,Boudet:2020eyr,Pigozzo:2020zft,Carneiro:2020uww}.} Although 
the recovery of the Bekenstein-Hawking entropy has been considered
as a way to fix its value, the  dependence of the
entropy calculation on $\gamma$ is controversial, and 
the value $\gamma\simeq 0.2375$ calculated thermodynamically is not universally accepted 
(see, for example, Refs.~\cite{Engle:2010kt,Bianchi:2012ui,Wong:2017pgl}).
One can argue that the semiclassical
thermodynamical properties can actually be recovered for any value of
$\gamma$ if one makes the appropriated assumptions. 
When introducing the notion of the horizon
into the quantum theory, it was shown that the entropy of large black
holes is independent of this parameter and, as it was believed, it is
only quantum gravity corrections to the entropy and temperature of
small black holes that depend on the Barbero-Immirzi
parameter. 

When treating $\gamma$ as a free parameter of the
theory, it is important to find ways to constraint its value 
from observations, if possible. The appearance of this parameter in the area and volume
spectra in LQG shows that it plays a role in determining the fundamental
length scale of space, since this parameter is used to count the size of the quantum of area in Planck units. If one could resolve distances of
around the Planck length, one would be able to fix the Barbero-Immirzi
parameter via  experiment. Once again, cosmology seems to be a window
of opportunity to access such scales, which are unreachable by any
terrestrial experiments.  As previously discussed, the quantum bounce
changes the scalar power spectrum by a correction term which depends
on the characteristic scale at the bounce. The characteristic scale is
the shortest scale (or largest wave number, namely $k_{B}$) that feels
the spacetime curvature during the bounce. Since this characteristic
scale is dependent on the value of the Barbero-Immirzi parameter, we have a unique
opportunity to constrain $\gamma$ through the observational limits on $k_B$ itself.

In  previous works (see, for instance.
Refs.~\cite{Zhu:2017jew,Benetti:2019kgw}), precise constraints on the
correction term in the power spectrum of LQC was obtained using
the recent CMB data. This observational analysis provided
constraints on the characteristic scale $k_{B}$.  This scale is a
function of the Barbero-Immirzi parameter and of the number of {\it
  e}-folds of expansion from the bounce until today. The suggestion of
considering the Barbero-Immirzi parameter as a free quantity was in
fact first considered in Ref.~\cite{Linsefors:2013cd}. In the present
paper, we, however, push that idea much further by taking advantage of
our analysis and results obtained in the first part of this
paper. Then, by using the observational constraints on $k_{B}$, we
also study their dependence on the Barbero-Immirzi parameter. This
will allow us to impose some precise limits on the value of $\gamma$
as a function of the {\it e}-folding number (or equivalently, to
obtain constraints on  the {\it e}-folds number as a function of
$\gamma$) depending on the primordial inflaton potential. 

In summary, in this paper we  keep investigating the intriguing
possibility that the quantum regime of the Universe in the context of
LQC could leave observable signals on CMB, going beyond
earlier works in at least two important aspects: (a) {}First, we
investigate the duration of inflation  for different potentials
including the Starobinsky and Higgs-like potential in addition to the
monomial ones, considering that the initial conditions  are uniquely
determined at the bounce and once we trace the dynamics far back in
the contracting phase. However, our methods are general enough to be
able to be extended to other forms of primordial potentials. {}For
practical purposes, we obtain the quantities at some point in the
contracting phase and where the kinetic energy of the inflaton starts
to dominate over that from the potential. {}From that point on, we can
forward the evolution up to the end of the accelerated inflationary
regime;  (b) second, we consider the {\it Barbero-Immirzi parameter}
as a free variable of the theory. {}From our analysis, we then obtain,
for the first time, a lower limit for this parameter by requiring  the
model to be consistent with the CMB observations. 
We have found constraints on the Barbero-Immirzi parameter in the context 
of the dressed metric approach and considering the BD vacuum as the 
initial condition for the perturbations in the contracting phase.  
Such choice provides basically the
same results as considering the fourth-order adiabatic vacuum state at the bounce as the
initial condition in the context of the dressed metric approach (for more details, see,
for instance, Ref.~\cite{Zhu:2017jew}).
{}Finally, we note that although some monomial potentials are already ruled out in the simple 
scenarios of cold inflation according to the Planck results, when radiation processes are present 
(most notably as is the case for these models when studied in the warm inflation context) all of 
these potentials can be shown to agree with the observations (see, e.g.,
Refs.~\cite{Benetti:2019kgw,Benetti:2016jhf}). 
Another reason for analyzing those models here is that they are well motivated
in the context of particle physics models in general.

This paper is organized as follows. In Sec.~\ref{Fundamentals}, we
briefly present the main equations of LQC, and also introduce
the potentials that we are going to consider in our analysis. In
Sec.~\ref{preanalysis}, we describe the method used in our analysis
and we also establish the way that the initial conditions can be
determined and that we are going to use in obtaining the subsequent
background dynamical evolution up to the end of inflation. In
Sec.~\ref{results}, we present the results obtained for the duration
of inflation in each model considered. The analytical results obtained
here are also compared with the numerical results obtained from
previous statistical analysis derived in Ref.~\cite{Barboza:2020jux}
and in other references. In Sec.~\ref{additional}, we discuss the
constraints on the value of the Barbero-Immirzi parameter from the
required number of inflationary {\it e}-folds in each model. In
Sec.~\ref{conclusion} we give our conclusions. One Appendix is
included to show some technical details.

\section{Theoretical Basis}
\label{Fundamentals}

In this section we briefly introduce the main  equations of LQC
and we also present the models we are going to consider in this
paper. 

As discussed in Ref.~\cite{Ashtekar:2011rm}, in LQC the spatial
geometry is described by the variable $\nu$ proportional to the
physical volume of a fiducial cubical cell, in place of the scale
factor $a$, i.e.,
 \begin{equation}\label{volume}
     \nu = - \frac{{\cal V}_{0}\,a^{3}\,m^{2}_{\rm Pl}}{2\pi \gamma},
 \end{equation}
where ${\cal V}_{0}$ is the comoving volume of the fiducial cell,
$m_{\rm Pl}\equiv 1/\sqrt{G}=1.22 \times 10^{19}\,$GeV is the Planck
mass, with $G$ the Newton's gravitational constant and $\gamma$ is the
Barbero-Immirzi parameter. Although $\gamma$ is actually a free
parameter of the theory, in the first part of our work we are going to
consider the value motivated from the calculation of the black hole
entropy in LQG, which is $\gamma \simeq
0.2375$~\cite{Meissner:2004ju},  which is the value considered in some other
studies~\cite{Engle:2010kt,Bianchi:2012ui}.

The {}Friedmann equation in LQC assumes the
form~\cite{Ashtekar:2011rm}
\begin{equation}\label{friedmann}
    \frac{1}{9}\,\left(\frac{\dot{\nu}}{\nu}\right)^{2} \equiv H^{2} =
    \frac{8 \pi}{3 m^{2}_{\rm Pl}} \rho \; \left(1 -
    \frac{\rho}{\rho_{cr}}\right),
\end{equation}
where
\begin{equation}
\rho_{cr} =  \frac{\sqrt{3}m_{\rm Pl}^4}{32 \pi^2 \gamma^3}
\label{rhoc}
\end{equation} 
is the critical density.  Through the modified {}Friedmann
equation~\eqref{friedmann}, we see explicitly the underlying quantum
geometric effects~\cite{Ashtekar:2011ni}, leading to a bounce in
replacement of the singularity  when $\rho = \rho_{cr}$. {}For $\rho
\ll \rho_{\rm cr}$ we recover GR as expected. 

The energy density $\rho$ in Eq.~\eqref{friedmann} respects the
usual conservation equation,
\begin{equation}\label{conservationeq}
 \dot{\rho} + 3H (\rho + p)  = 0,
\end{equation}
where $p$ is the pressure density.

In models of a single scalar field $\phi$ with a potential $V(\phi)$,
in the {}Friedmann-Lema\^itre-Robertson-Walker (FLRW)
background, the effective equation of motion for $\phi$, the inflaton,
is simply
\begin{equation}
 \ddot{\phi} + 3H \dot{\phi} + V_{,\phi} = 0,
\label{eom}
\end{equation}
where $V_{,\phi} \equiv dV(\phi)/d\phi$ is the derivative of the
inflaton's potential.

In the following, we are going to  present the models we are going to
consider.

\subsection{Models}
\label{modelsP}
The classes of inflationary models we are going to consider are
described by the potentials below.

\subsubsection{Power-law monomial potentials}
\label{chaotic}

In this class of models, the potential is given by 
\begin{equation}\label{powerlawV}
V= \frac{V_0}{2n} \left( \frac{\phi}{m_{\rm Pl}}\right)^{2n},
\end{equation}
where $n$ is some power. In this paper we will compare our results
with available ones for the cases for the quadratic, quartic and
sextic forms of the potential (corresponding to the powers $n=1,\,2\;
{\rm and}\; 3$, respectively). The model given by
Eq.~(\ref{powerlawV}) covers the class of inflationary models
corresponding to large-field models~\cite{Lyth:2009zz}. 

The new data release from BICEP/Keck \cite{BICEP:2021xfz} strengthened
the bounds on the tensor-to-scalar ratio $r$, putting severe
constraints in  the full class of monomial potentials, showing that
they are  disfavored in the context of standard inflation. However,
these potentials, besides being well motivated from a particle
physics point of view, can still be perfectly allowed in other
frameworks, like in the context of warm inflation for example (see,
e.g., Ref.~\cite{Benetti:2016jhf} and references therein). Since our
current analysis can be extended to such frameworks, we consider it
important to address this class of potentials here. 

\subsubsection{The Higgs-like symmetry breaking  potential}

The Higgs-like symmetry breaking  potential is given by the following
expression,
\begin{equation}\label{higgsV}
V=\frac{V_{0}}{4\,m_{\rm Pl}^{4}} (\phi^{2} - v^{2})^{2},
\end{equation}
where $v$ denotes the vacuum expectation value (VEV) of the
field.  The Higgs-like symmetry breaking  potential is another well
motivated model from particle physics. Besides that, it can
represent either a small-field inflation model, if inflation starts
(and ends) at the small field part of the potential (i.e., for $|\phi|
< |v|$), or be a large field model, if inflation happens in the large
field region of the potential ($|\phi| > |v|$). In all our analyses
with this potential, we  have explicitly distinguished these two
possibilities and produced results for both of them and by also
considering different values for the VEV $v$.

\subsubsection{The Starobinsky  potential}

The Starobinsky model~\cite{Starobinsky:1980te}  is an example of a
limiting case of more general modified gravity theories. When
expressed in the Einstein frame, it represents a potential which can
be written as
\begin{equation}\label{starobinskyV}
V= V_0 \left( 1-e^{-\sqrt{\frac{16\pi}{3}}\frac{\phi}{m_{\rm
      Pl}}}\right)^{2}.
\end{equation}
The inflation results derived from the model Eq.~(\ref{starobinskyV})
agree quite well with the observational data for the tensor-to-scalar
ratio and spectral tilt~\cite{Planck:2018jri}.  {}For that reason, it
is a popular form of potential in inflation studies.

Note that in Eqs.~(\ref{powerlawV})
(\ref{starobinskyV}), the scale $V_{0}$  is fixed differently
depending on the potential. Its value is determined by the amplitude
of the CMB scalar spectrum. Details about its evaluation for the
three forms of potentials considered here are giving in the
Appendix for completeness.

\section{Background Dynamics in LQC}
\label{LQCphases}
 
The dynamics in the LQC models considered here is assumed to start in
the contracting phase sufficiently before the bounce and as originally
assumed to be the appropriate moment for setting the initial
conditions~\cite{Linsefors:2013cd}.  Assuming that the initial
conditions are set in the contracting phase, when the inflaton field
is in the oscillating regime, then,  considering a generic inflaton
potential of the form $V\propto \phi^m$, $m>0$, it then follows that
the inflaton field amplitude is expected to evolve as a function of
the scale factor $a(t)$ as~\cite{Turner:1983he}  $\phi\propto
a(t)^{-6/(m+2)}$.  Thus, in the oscillating regime we have that
$V\propto a(t)^{-6m/(m+2)}$, while the inflaton's kinetic energy will
behave like $\dot \phi^2 \propto a(t)^{-6}$. Thus, for any finite
value for the exponent $m$, with the oscillating phase lasting long
enough in the contracting phase, the  kinetic energy of the inflaton
field will necessarily come to dominate at the bounce. As we are going
to show in this section, with the kinetic energy dominating at the
bounce, the value of the inflaton field, $\phi_{B}$, will then be
uniquely determined at this moment. This result is independent on the
details of the contracting phase, provided that the evolution starts
sufficiently back in the contracting phase and in the oscillatory
regime for the inflaton. The obtained value for $\phi_B$ for each
model provides the initial conditions for the further evolution of the
system and which can be carried out up to the end of the accelerated
inflationary regime.

In LQC models in which the evolution of the inflaton field is
dominated by its kinetic energy at the quantum bounce, a slow-roll
inflation phase is practically always reached as demonstrated in many
previous
papers~\cite{Ashtekar:2011rm,Linsefors:2013cd,Bolliet:2017czc,Zhu:2017jew,Graef:2018ulg,Barboza:2020jux}. Our
goal here will be to estimate the number of {\it e}-folds of expansion
in two regimes: In the preinflationary one, which includes the
instant of the bounce until the beginning of slow-roll inflation. We
denote this number of {\it e}-folds of expansion as $N_{\rm
  pre}$. Then, the  {\it e}-folds from the beginning to the end of
inflation, which we call $N_{\rm infl}$, is also determined. In order
to achieve this, let us first begin by analyzing the background
dynamics in our scenario starting from the bounce phase. We will
proceed without assuming a specific model for the contracting phase. 

Immediately before and after the bounce, assumed to happen at a time
instant $t_B$, if the energy density is mostly dominated by kinetic
energy as we are going to consider and discussed above, we have a
phase of {\it superdeflation} (for $t< t_B$) and  {\it superinflation}
(for $t>t_B$). These are very short phases, which start close to the
bounce instant (when $H = 0$, i.e., $\rho=\rho_{\textrm{cr}}$). They
start (at $t<t_B$) and end (at $t>t_B$) when $\dot{H} = 0$. The
conditions for superdeflation/superinflation are
\begin{align}
H^{2} \gg |V_{,\phi}|, \quad   \dot\phi^{2}/2 \gg V(\phi).
\label{superinfl}
\end{align}
Right after the superinflation phase, in the postbounce phase, the
kinetic energy quickly decreases as $\dot{\phi}^{2} \propto 1/a^{6}$,
while the potential energy density $V(\phi)$ only changes slowly. The
inflaton dynamics after the bounce and throughout the preinflationary
phase is just monotonic~\cite{Zhu:2017jew}. After a given moment, the
potential energy of the inflaton will eventually dominate the energy
content of the Universe and the standard slow roll inflationary phase
will set in.  Before the beginning of slow roll, the quantum
corrections to the {}Friedmann equation are negligible, and the
cosmological equations reduce to the usual ones of GR, at which
point we then have that
\begin{align}
\rho_{cr} \ll \rho, \quad  V(\phi) \gg  \dot\phi^{2}/2.
\label{superinfl2}
\end{align}
More specifically, the evolution of the Universe for $t \geq t_B$ and
prior to reheating (end of inflation) is divided into three different
phases: {\it the bouncing}, {\it the transition} and {\it slow-roll
  inflation}~\cite{Zhu:2017jew}. Next, we describe each of these
phases. 

\subsection{Kinetic dominated regime}

This phase is dominated by the kinetic energy of the inflaton, with
$\dot{\phi}^2/2 \gg V(\phi)$. In Eq.~\eqref{eom}, neglecting the
derivative of the potential~$V_{,\phi}$, we have
\begin{equation}\label{moviment_SR}
    \Ddot{\phi} + 3 H \dot{\phi} \approx 0,
\end{equation}
whose solution is given by 
\begin{equation}\label{dotphi}
    \dot{\phi}(t)= \pm \sqrt{2 \rho_{\rm cr}} \left[\frac{a_{\rm
          B}}{a(t)}\right]^{3},
\end{equation}
where $a_{\rm B}$ is the scalar factor at the bounce. Substituting the
solution (\ref{dotphi}) in Eq.~\eqref{friedmann}, we obtain
\begin{equation}\label{factor_evol}
    a(t) = a_{\rm B} \left[1  +\frac{24 \pi \rho_{\rm cr}}{m_{\rm
          Pl}^{4}}\,\frac{(t-t_{\rm B})^2}{t_{\rm Pl}^2}\right]^{1/6},
\end{equation}
where $a_{\rm B}\equiv a(t_{\rm B})$ and $t_{\rm B}$ is the bounce
instant. Equation~(\ref{factor_evol}) gives the expression for the
scale factor in the quantum regime of the Universe. In the above
equation, $t_{\rm Pl}=1/m_{\rm Pl}$ is the Planck time.

With the analytical solution for $a(t)$, from \eqref{dotphi} one finds
\begin{equation}\label{critical_movim}
    \dot{\phi}(t) = \pm \frac{\sqrt{2\rho_{\rm cr}}}{\left[1+ \frac{24
          \pi \rho_{\rm cr}}{m_{\rm Pl}^4}\,(t-t_{\rm B})^{2}/t_{\rm
          Pl}^{2}\right]^{1/2}},
\end{equation}
and
\begin{equation}\label{mov_evol}
\phi(t) = \phi_{\rm B} \pm \frac{m_{\rm Pl}}{2\sqrt{3
    \pi}}\,\operatorname{arcsinh}{\left[\sqrt{\frac{24 \pi \rho_{\rm
          cr}}{m_{\rm Pl}^4}}\,\frac{(t-t_{\rm B})}{t_{\rm
        Pl}}\right]}.
\end{equation}

\begin{center}
\begin{figure}[!htb]
\subfigure[]{\includegraphics[width=7.3cm]{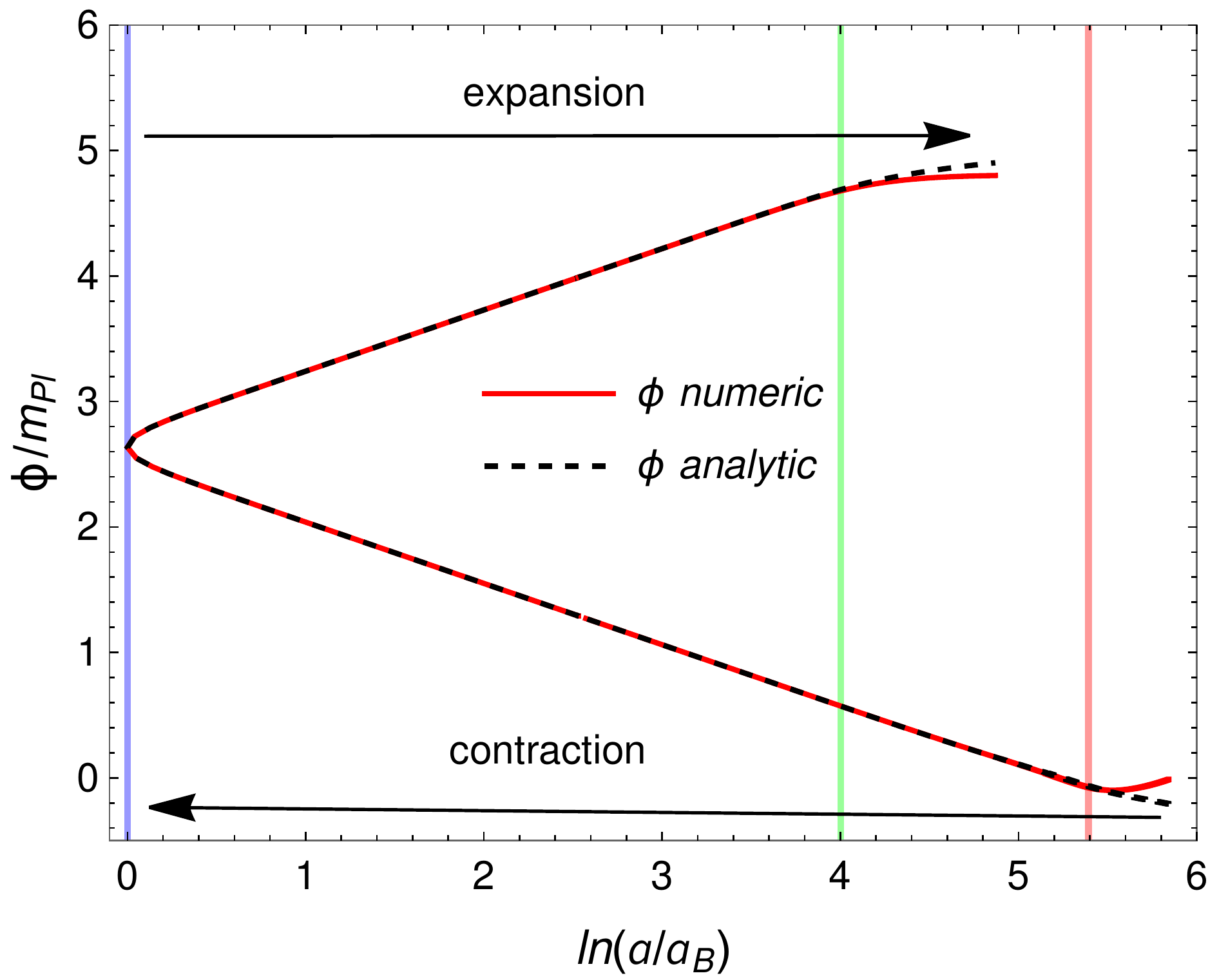}}
\subfigure[]{\includegraphics[width=7.3cm]{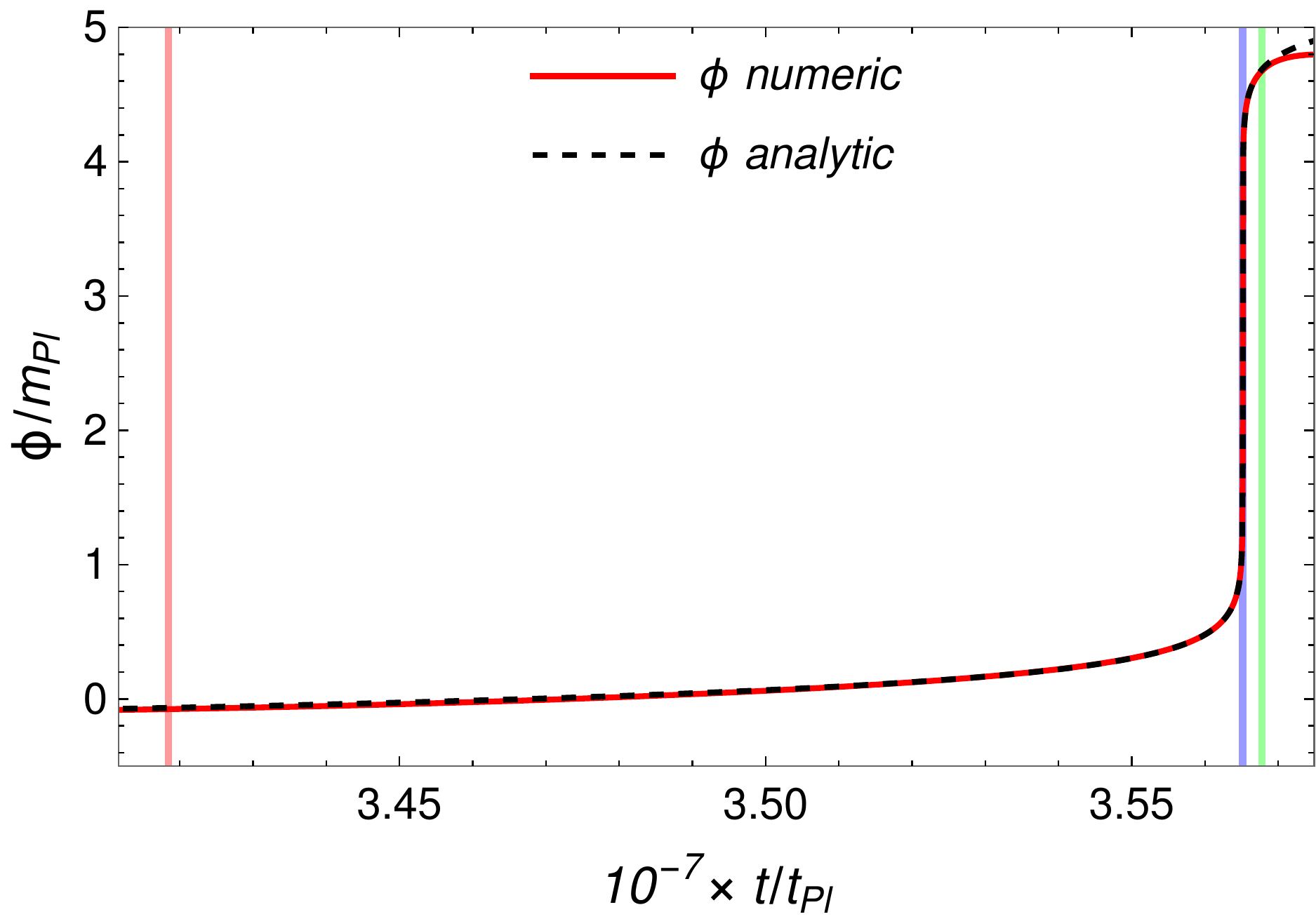} }
 \caption{Comparison of the numerical and analytical results for the
   inflaton amplitude $\phi$.  A quadratic power-law potential was
   considered and initial conditions (set in the contracting phase)
   such that around $147$ {\it e}-folds of inflation are generated (see
   text).  The vertical red and green strips indicate the start and
   end of the kinetic energy dominated regime, $\dot{\phi}^2/2 >
   V(\phi)$. The vertical blue strip indicates the bounce instant
   $t_B$. Evolution is shown in terms of both the number of {\it e}-folds
   [panel (a)] and in terms of the physical time [panel (b)].}
 \label{fig1}
 \end{figure}
\end{center}

It is important to notice that even though the result for the inflaton
amplitude Eq.~(\ref{mov_evol}) is derived close to the bounce and
where the inflaton potential is negligible, its validity still extends
quite well for a long time interval both before and after the
bounce. This is illustrated in {}Fig.~\ref{fig1}, where we show the
numerical evolution for the inflaton amplitude $\phi(t)$ and compare
it to the analytical result given by Eq.~(\ref{mov_evol}). {}For
illustrative purposes, we have considered the quadratic power-law
potential, Eq.~(\ref{powerlawV}), with $n=1$, but the results remain
qualitatively similar when considering other potentials. The initial
conditions were considered deep in the contracting phase and such that
around $147$ {\it e}-folds of inflation would be produced. This value of
{\it e}-folds of inflation was chosen since it is within what is expected
for this potential from previous statistical analysis for this form of
potential (see, e.g., Refs.~\cite{Linsefors:2013cd,Barboza:2020jux}
for details). The transition time in the contracting and expanding
phases, $t^{-}_{tr}$ and $t_{tr}^+$, respectively, defined by
$\dot{\phi}^2(t_{tr}^{\pm})/2 = V(\phi(t_{tr}^{\pm}))$, are marked by
the red and green vertical strips, respectively.  The bounce is marked
by the blue vertical strip. Note that the analytical result for
$\phi(t)$ agrees well with the numerical solution obtained when
evolving Eq.~(\ref{eom}) with the Hubble parameter in LQC given by
Eq.~(\ref{friedmann}) in the whole region $t^{-}_{tr} \lesssim t
\lesssim t_{tr}^+$.

\subsection{Setting the initial conditions}
\label{preanalysis}

Let us now describe the process of determining the appropriate initial
conditions for the evolution of the system and which will determine
the inflaton amplitude at the bounce instant, $\phi_{\rm B}$. Having
$\phi_B$ is important because we can  straightforwardly relate all
relevant postbounce quantities for our analysis with it, as we are
going to show below.

{}First, one notices that for a fluid with equation of state $w$, we
can  write the Hubble parameter in LQC as
\begin{equation}
H_{\rm LQC}(t)= \frac{ \frac{4 \pi \rho_{\rm cr}}{m_{\rm Pl}^4} (1+w)
  \frac{t-t_B}{t_{\rm Pl}}} {\left[ 1+ \frac{6 \pi \rho_{\rm
        cr}}{m_{\rm Pl}^4} (1+w)^2 \frac{(t-t_B)^2}{ t_{\rm Pl}^2}
    \right] } m_{\rm Pl},
\label{HLQCw}
\end{equation}
which comes straightforwardly when we generalize the scale factor
Eq.~(\ref{factor_evol}) to an arbitrary constant equation of state
$w$. The results derived next hold far from the bounce, where
the LQC effects are negligible.  {}Far from the bounce, when the quantum effects can be
neglected, {\it i.e.}, $ |t-t_B| \gg t_{\rm Pl}$, Eq.~(\ref{HLQCw})
gives the usual expression  obtained in GR for a fluid with constant
equation of state $w$,
\begin{equation}
H \simeq \frac{2}{3(1+w) (t-t_B)}.
\label{Hw}
\end{equation}
In the following we will assume that the equation of state in
Eq.~(\ref{Hw}) can be extended for that of the inflaton field,
{\it i.e.},
\begin{eqnarray}
w &=& \frac{\dot{\phi}^2/2 - V(\phi)}{\dot{\phi}^2/2 + V(\phi)}.
\label{wphi}
\end{eqnarray}
{}Following also a notation analogous to the one used by the authors
of Ref.~\cite{Ashtekar:2011rm} and defining the ratio of the potential
energy  to kinetic energy, $\alpha \equiv V/(\dot \phi^2/2)$, then
the Hubble parameter far from the bounce is
approximated as
\begin{equation}
H \simeq \frac{1+\alpha}{3 (t-t_B)},
\label{Halpha}
\end{equation}
where we have used that $w=(1-\alpha)/(1+\alpha)$ in Eq.~(\ref{Hw}).  
In principle the
expression (\ref{Halpha}) might be considered a too rough
approximation for the Hubble parameter. But as already noticed before,
e.g., from the results shown in {}Fig.~\ref{fig1}, if we look at the
dynamics in the contracting phase, for $t_{tr}^- < t < t_B$, it tends
to be much slower  than in the expanding phase, for $t_B < t < t_{tr}^+$,
which gives us hope that we can eventually match the results at some
point in the contracting phase with some value of $\alpha$ and then
evolve the system forward to the bounce instant. As suggested from
the results shown in {}Fig.~\ref{fig1} and the above discussion, the
appropriate moment to perform such matching is well before the bounce, 
while in the contracting phase, but where the quantum effects are
still negligible and we can use Eq.~(\ref{Halpha}), but after the transition
time $t_{tr}^-$, where we can still apply Eqs.~(\ref{critical_movim}) and
(\ref{mov_evol}) with good accuracy. Our results to be
shown in Sec.~\ref{results} indeed indicate that this simple
strategy is a sound one.

Our approximation then consists of considering the dynamics at some
point in the contracting phase, after the transition time $t_{tr}^-$,
but still well before the bounce, such that the quantum effects are
still negligible. We then look at some instant $t_\alpha$ in the
contracting phase given by $t_{tr}^- < t_\alpha \ll t_B$, and where
the Hubble parameter can be approximated by Eq.~(\ref{Halpha}).
Taking the time derivative of Eq.~(\ref{Halpha}) and equating it to
$-4 \pi \dot{\phi}^2/m_{\rm Pl}^2$, which is valid in the regime we
are considering ({\it i.e.}, far from the bounce and where the quantum
effects are still negligible), we find a direct relation between the
potential and its field derivative, $V'$, as given by
\begin{equation}
\frac{V(\phi_\alpha)}{V'(\phi_\alpha)} = \frac{\sqrt{1+\alpha}}{4
  \sqrt{3 \pi}}   m_{\rm Pl}.
\label{Vratio}
\end{equation}
To obtain Eq.~(\ref{Vratio}),  we have used the inflaton's equation of
motion Eq.~(\ref{eom}) along also with Eq. (\ref{Halpha}) to eliminate
the explicit time dependence in favor of the Hubble parameter and
finally that\footnote{Note the choice of minus sign for the Hubble
  parameter is because we are considering the dynamics in the
  contracting phase, hence, $H<0$.} 
\begin{equation}
H=-\sqrt{ \frac{8 \pi}{3 m_{\rm Pl}^2} \frac{(1+ \alpha)}{\alpha} V},
\label{HalphaV}
\end{equation}
which follows when using that $\dot \phi^2/2 \equiv V/\alpha$.  Our
approximation to estimate $\phi_B$ now consists of considering that we
can take an ``average" value for $\alpha$ and approximate it as a
constant $\bar \alpha$.  In this case, for a given value of $\bar
\alpha $ within the range $(0,1)$ we can readily estimate the inflaton
amplitude $\phi_\alpha\equiv \phi(t_\alpha)$, for any given potential,
when using Eq.~(\ref{Vratio}), and also the instant of time $t_\alpha$
when using Eqs.~(\ref{Halpha}) and (\ref{HalphaV}) for $\alpha \to
\bar \alpha$,
\begin{equation}
t_\alpha - t_B = - \frac{1+\bar \alpha}{3}  \sqrt{ \frac{3 m_{\rm
      Pl}^2 \bar \alpha}{8 \pi (1+\bar \alpha) V(\phi_\alpha)}}.
\label{talpha}
\end{equation}
Here we fix the value for $\bar \alpha$ such that the dynamics will
match the one obtained for each potential and which results are
available for the potentials we are analyzing. We note that this
strategy is analogous to the one adopted, e.g., by the authors of
Ref.~\cite{Ashtekar:2011rm}, where the constant value for $\bar
\alpha$ was fixed by matching their numerical results for $\phi$, but
in the postbounce regime instead.  The apparent arbitrariness in
having to choose a different value of $\bar \alpha$  for different
initial conditions has also been dealt with in
Ref.~\cite{Ashtekar:2011rm} by fixing $\bar \alpha$ for one initial
condition and then using the {\it same} value for all other
cases. Thus, there is only one value of $\bar \alpha$ fixed once and
for all. As shown in Ref.~\cite{Ashtekar:2011rm}, this simple
approach was good enough to reproduce their numerical results in the
postbounce phase when considering different initial conditions taken
at the bounce instant.  Here we avoid applying this to the postbounce
regime and choose to consider its application in the contracting phase
instead. This is justified because, as seen from {}Fig.~\ref{fig1}(b),
the postbounce dynamics lasting from the bounce up to the transition
time $t_{tr}^+$ is much shorter than the one  lasting from $t_{tr}^-$
in the contracting phase until the bounce time $t_B$.  The ratio of
energies $\alpha$ changes much faster in the expanding phase than in
the contracting one.  Trying to fix $\alpha$ to match the numerical
results in the expanding phase and evolving back to the bounce instant
$t_B$ to obtain $\phi_B$ for example, thus implies  requiring a much
higher accuracy than the one we can achieve doing the same procedure
in the contracting phase. Motivated also by the results obtained by
the authors of Ref.~\cite{Ashtekar:2011rm}, we will fix the value of
$\bar \alpha$ such that our analytical results will match the
statistical analysis considered for example in
Ref.~\cite{Barboza:2020jux}. {}Furthermore, we also show that we only
need to fix the value of $\bar \alpha$ once for one primordial
inflaton  potential. This same value can then be used for all other
potentials.  We will see that this simple strategy will produce
results with good accuracy as confirmed by the results shown in
Sec.~\ref{results}.
\subsection{The postbounce transition time and the inflaton amplitude}

Having both $\phi_\alpha$ and the instant $t_\alpha$ for a given value
of $\bar \alpha$ by following the above procedure in the contracting
phase, we then can use Eq.~(\ref{mov_evol}) to obtain the inflaton
amplitude at the bounce time, $\phi_B$. Given the value for $\phi_B$,
we can estimate the number of {\it e}-folds of inflation. {}First, the
inflaton value at the transition time $t_{tr}^+$ in the postbounce
regime is determined by using Eq.~(\ref{mov_evol}) again, 
\begin{equation}
 \phi(t_{tr}^+) =\,\phi_{ B}\,+ \,\frac{m_{\rm Pl}}{2 \sqrt{3 \pi}} \,
 \operatorname{arcsinh}\left( \sqrt{\frac{24 \pi \rho_{\rm cr}}{m_{\rm
       Pl}^4}}\, \frac{t_{tr}^+ - t_B}{t_{\rm Pl}}\right).
\label{phic_ln}
\end{equation}
We can now consider that at the transition time we have
that\footnote{Note that we could have this equation for both signs
  positive or negative for $\dot \phi$. Throughout this paper, we work
  with the convention of adopting the positive sign for $\dot \phi$,
  thus also considering the positive sign in Eq.~(\ref{mov_evol}) when
  writing it for $t=t_{tr}^+$ in Eq.~(\ref{phic_ln}). This implies
  that the field is always moving from the left to the right side of
  the potential. {}For the power law and Higgs potentials this choice
  does not lead to any ambiguity since the potential is symmetric and
  for any choice of the sign for $\dot \phi$ the field is always
  climbing the potential when starting the initial conditions deep
  inside the contracting phase and always close to the minimum of the
  potential. The Starobinsky potential is asymmetric, but for the
  standard form, Eq.~(\ref{starobinskyV}), inflation only happens
  along the flat region, which resides in the right-hand side of the
  potential.} 
\begin{equation}
\dot{\phi}(t_{tr}^+) = \sqrt{2 V(\phi(t_{tr}^+))}.
\label{dotphiV}
\end{equation}
By using the time derivative of $\phi(t)$, Eq.~(\ref{critical_movim})
at $t_{tr}^+$ and  substituting Eq.~(\ref{phic_ln}) in
Eq.~(\ref{dotphiV}), we can then numerically\footnote{In fact exact
  analytical expressions for both $t_{tr}^+-t_B$ and $\phi(t_{tr}^+)$
  can be obtained from these equations by approximating them by
  considering that $t_{tr}^+-t_B \gg t_{\rm Pl}$ (see, e.g.,
  Ref.~\cite{Zhu:2017jew} for details in the cases of the quadratic
  power law and Starobinsky potentials). The solution is in general
  expressed in terms of Lambert functions. Here we simply choose to
  directly numerically solve Eq.~(\ref{dotphiV}), which can  in
  principle be done for any potential in general.} solve
Eq.~(\ref{dotphiV}) for the time interval $t_{tr}^+-t_B$.  This result
then also allows us to obtain  $\phi(t_{tr}^+)$ when substituting the
solution for  $t_{tr}^+-t_B$ back in Eq.~(\ref{phic_ln}).

\subsection{Beginning of the slow-roll inflationary phase}

When the slow-roll phase starts at some time $t_{\rm i} > t_{tr}^+$,
the Universe is already far from the quantum regime. The potential
energy starts to dominate over the kinetic energy giving rise to the
inflationary regime.  In the following, we use the index ``i"
to denote the quantities at the beginning of inflation, which
corresponds to the moment when the Universe starts accelerating and
the equation of state satisfies $w\leq -1/3$. In order to obtain the
quantities in this moment, we can use the expansion for $\phi(t)$
which is valid for $t \simeq t_{\rm i}$,
\begin{equation}\label{phi_initial}
    \phi_{\rm i} \simeq \phi_{\rm tr} + \dot{\phi}_{\rm
      tr}\,t_{tr}^+\,\ln{\frac{t_{\rm i}}{t_{tr}^+}},
\end{equation}
where $\phi_{\rm tr}\equiv \phi(t_{tr}^+)$, $\dot{\phi}_{\rm tr}
\equiv \dot{\phi}(t_{tr}^+)$ and, without loss of generality, we are
setting from this point on that $t_B=0$.  Likewise for  $\dot
\phi(t_{\rm i})$, we have that
\begin{equation}
    \dot{\phi}_{\rm i} \simeq \frac{t_{tr}^+}{t_{\rm i}} \,
    \dot{\phi}_{\rm tr}.
\end{equation}
Thus, 
\begin{equation}
    V(\phi_{\rm i}) \simeq V(\phi_{\rm tr}) + V_{,\phi} \, (\phi_{\rm
      tr}) \, t_{tr}^+ \,  \dot{\phi}_{\rm tr} \ln{\frac{t_{\rm
          i}}{t_{ tr}^+}}.
\label{potVti}
\end{equation}
Since the accelerated regime, $\ddot a >0$, starts at $w=-1/3$ for the
equation of state, then
\begin{equation}
 \dot{\phi}_{\rm i}^{2}=V(\phi_{\rm i}).
\label{dotphiVti}
\end{equation}
Using Eqs.~(\ref{dotphiVti}) and (\ref{potVti}) and knowing $t_{tr}^+$
and $\phi_{\rm tr}$ obtained from the  previous step, we can now
numerically solve\footnote{Again, it can be found general analytical
  solutions for Eq.~(\ref{dotphiVti}) which are expressed also in
  terms of Lambert functions~\cite{Zhu:2017jew}, but for practical
  purposes we just opt to numerically solve Eq.~(\ref{dotphiVti}).}
Eq.~(\ref{dotphiVti}) for the initial time $t_i$, which will then
determine $\phi_i$ from Eq.~(\ref{phi_initial}).

\subsection{Number of {\it e}-folds $N_{\rm pre}$ and $N_{\rm infl}$}
\label{analysisa}

The number of {\it e}-folds of expansion is defined as
\begin{equation}
   N \equiv \ln\left(\frac{a_{\rm end}}{a_{\rm init} }\right),
\end{equation}
where $a_{\rm init}$ and $a_{\rm end}$ denote, respectively, the
scale factors at the beginning and at the end of the
corresponding period.

Let us first present the results for the preinflationary phase,
i.e., the expansion lasting from the bounce time to the start of the
inflationary phase.  The number of {\it e}-folds for the
preinflationary phase is denoted by $N_{\rm pre}$, with $N_{\rm pre}
\equiv \ln (a_{\rm i}/a_{\rm B})$, where $a_{\rm B}$ and $a_{\rm i}$
are the scale factors at the bounce and at the beginning of inflation,
respectively. According to what we have done in the previous sections,
we can write $a_{\rm i}$ as~\cite{Zhu:2017jew}
\begin{equation}\label{aitr}
   a_{\rm i} \simeq a_{\rm tr}\left( 1 +  t_{tr}^+\,H_{\rm tr}
   \ln{\frac{t_{\rm i}}{t_{tr}^+}}\right),
\end{equation}
where $H_{\rm tr}$ in Eq.~\eqref{aitr} is obtained from the
{}Friedmann equation  considering $\rho_{\rm tr}=\dot{\phi}_{\rm
  tr}^{2}/2 + V (\phi_{\rm tr})$.  Therefore, the number of {\it
  e}-folds of expansion in the preinflationary phase can be written
as
\begin{eqnarray}
   N_{\rm pre} &=& \ln \left(\frac{a_{\rm i}}{a_{\rm B}}\right) =\ln
   \left(\frac{a_{\rm tr}}{a_{\rm B}}\right) +\ln \left(\frac{a_{\rm
       i}}{a_{\rm tr}}\right) \nonumber \\ &\simeq &  \frac{1}{6} \ln
   \left(1 + \frac{24 \pi \rho_{\rm cr}}{m_{\rm Pl}^4}
   \frac{(t_{tr}^+)^2}{t_{\rm Pl}^{2}}\right) + \ln \left(1 +
   t_{tr}^+\,H_{\rm tr} \ln{\frac{t_{\rm i}}{t_{tr}^+}}\right),
   \nonumber \\
\label{Npre}
\end{eqnarray}
where we have used Eq.~\eqref{factor_evol} at $t=t_{tr}^+$ and
Eq.~(\ref{aitr}).

The number of the {\it e}-folds of expansion during the inflationary
phase, $N_{\rm infl}$, is defined as,
\begin{equation}\label{N_folds}
    N_{\rm infl}(\phi) \equiv \ln{\left(\frac{a_{\rm end}}{a_{\rm
          i}}\right)} \approx  \frac{8 \pi}{m^{2}_{\rm
        Pl}}\int^{\phi_{\rm i}}_{\phi_{\rm end}}
    \left(\frac{V}{V^{\prime}}\right)\,d\phi,
\end{equation}
where in the last term in the above equation we have used the
slow-roll approximations, valid during inflation, $\dot \phi \simeq
-V_{,\phi}/(3H)$  and $H^2 \simeq 8\pi V/(3 m_{\rm Pl}^2)$.  The total
number of {\it e}-folds lasting from the bounce until the end of
inflation is then $N_{\rm total} = N_{\rm pre}  + N_{\rm infl}$.  In
Eq.~(\ref{N_folds}), $\phi_{\rm i}$ is given by
Eq.~(\ref{phi_initial})  and $\phi_{\rm end}$, the scalar field at the
end of the inflationary phase, is obtained by the slow-roll
coefficient $\epsilon = - \dot{H}/{H^2}$ when it is set to one
(indicating the end of the accelerated regime).  Thus, from $\epsilon
= - \dot{H}/{H^2} \simeq (V_{,\phi}m_{\rm Pl}/V)^{2}/(16\pi) = 1$,
$\phi_{\rm end}$ can be readily obtained for each of the potential
forms we are considering. The results are explicitly given below.

\subsubsection{The power-law monomial potentials}

{}From Eq.~(\ref{N_folds}),  $N_{\rm infl}$ in the case of monomial
potentials is given by
\begin{eqnarray}\label{ninf_powerlaw}
    N_{\rm infl}&=&\frac{2 \pi}{n\,m^{2}_{\rm Pl}} \left(\phi_{\rm
      i}^{2} - \phi^{2}_{\rm end}\right),
\end{eqnarray}
with  $\phi_{\rm end}$  given by
\begin{equation}\label{phiend}
  \phi_{\rm end}^{2} = \frac{n^{2}}{4 \pi} m^{2}_{\rm Pl}.  
\end{equation}

\subsubsection{The Higgs-like symmetry breaking potential}

{}For the  Higgs-like potential we have that
\begin{equation}\label{ninf_higgs}
   N_{\rm infl} = \frac{2 \pi}{m^{2}_{\rm Pl}} \left[\frac{\phi_{\rm
         i}^{2} - \phi_{\rm end}^{2}}{2} - v^{2} \ln{\frac{\phi_{\rm
           i}}{\phi_{\rm end}}}\right]
\end{equation}
and
\begin{equation}\label{phiend_Higgs}
  \phi_{\rm end} = \pm \sqrt{v^{2} + \frac{m^{2}_{\rm Pl}}{2
      \pi}\,\pm\,\frac{m^{2}_{\rm Pl}}{2 \pi} \sqrt{1 + \frac{4 \pi
        v^{2}}{m^{2}_{\rm Pl}}}},
\end{equation}
where the signs positive and negative correspond to the large and
small-field cases, respectively.

\subsubsection{The Starobinsky potential}

{}For the Starobinsky potential we obtain that
\begin{equation}\label{ninf_starobinsky}
N_{\rm infl} = \frac{3}{4} \left( e^{ \sqrt{ \frac{16\pi}{3} } \frac{
    \phi_{\rm i} }{ m_{\rm Pl} } } - e^{ \sqrt{ \frac{16\pi}{3} }
  \frac{ \phi_{\rm end} }{ m_{\rm Pl} } } \right)  +
\frac{\sqrt{3\pi}}{m_{\rm Pl}} \left(\phi_{\rm i} - \phi_{\rm
  end}\right),  
\end{equation} 
where
\begin{equation}\label{phiend_starobinsky}
\phi_{\rm end} = m_{\rm
  Pl}\sqrt{\frac{3}{16\pi}}\ln{\left(1+\frac{2}{\sqrt{3}}\right)} .   
\end{equation}

Having derived and collected all relevant equations,  we are now in
condition to present our results in the following section.

\section{Results}
\label{results}

\begin{center}
\begin{figure}[!htpb]
\subfigure[]{\includegraphics[width=7.3cm]{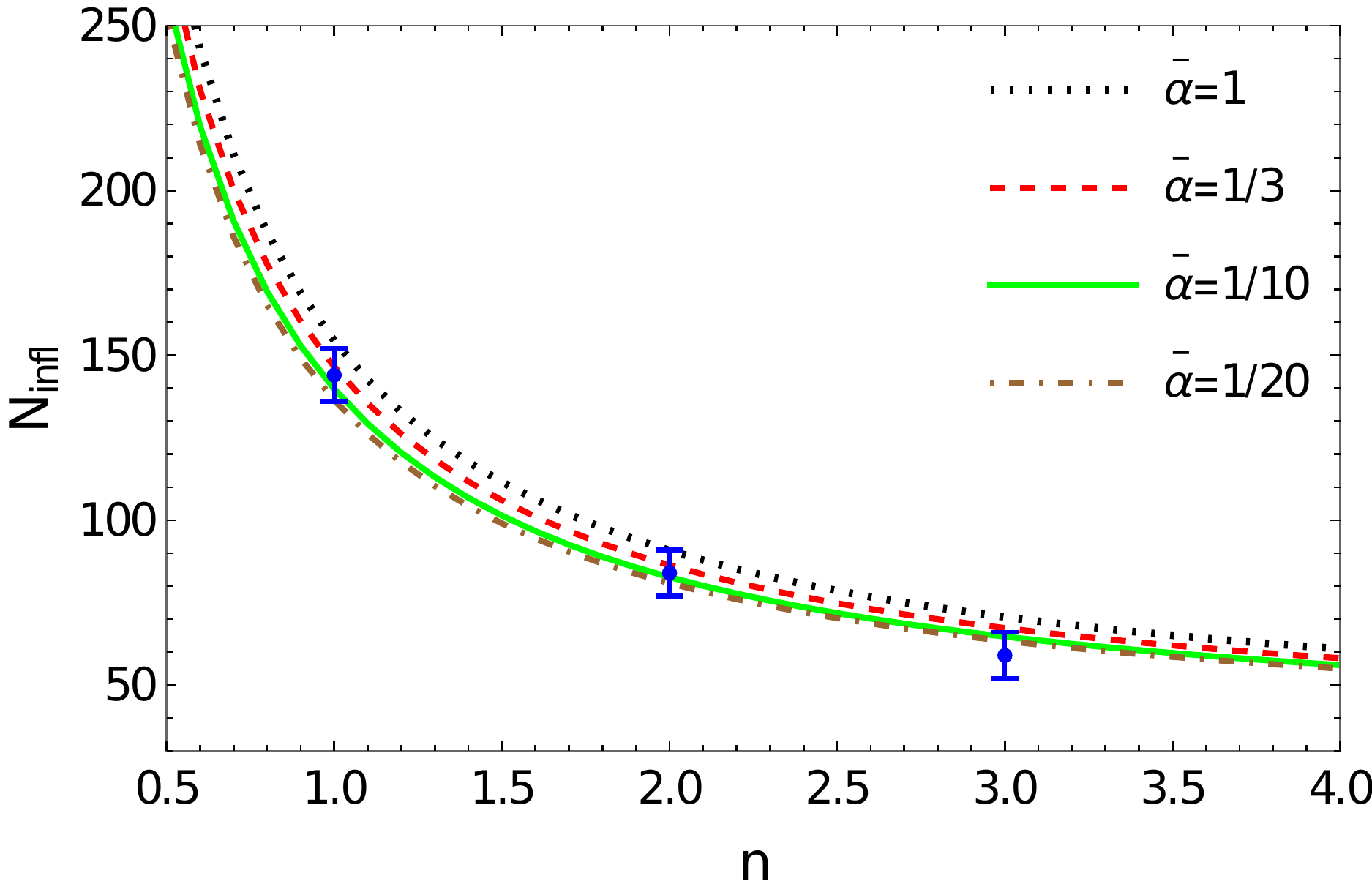}}
\subfigure[]{\includegraphics[width=7.3cm]{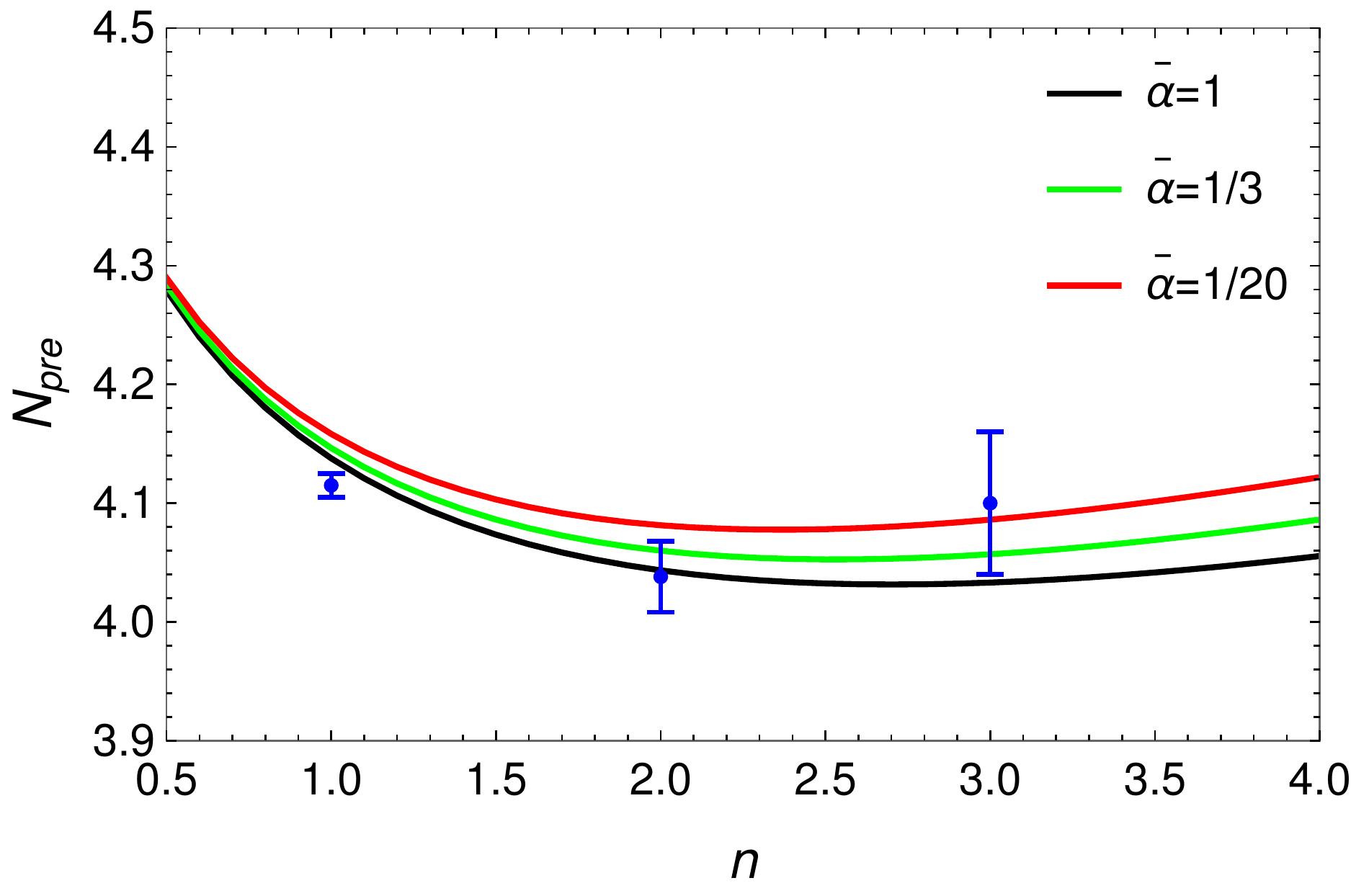} }
 \caption{The numerical results for the number of {\it e}-folds of inflation
   [panel (a)] and for the number of {\it e}-folds for the preinflationary
   regime lasting from the bounce until the beginning of inflation
   [panel (b)] obtained through the method described in
   Sec.~\ref{analysisa} when applied to the monomial power-law
   potential Eq.~(\ref{powerlawV}).  The data points show the
   results obtained in Ref.~\cite{Barboza:2020jux}.}
 \label{fig2}
 \end{figure}
\end{center}

\begin{center}
\begin{figure}[!htpb]
\subfigure[$|\phi| > |v|$]{\includegraphics[width=7.3cm]{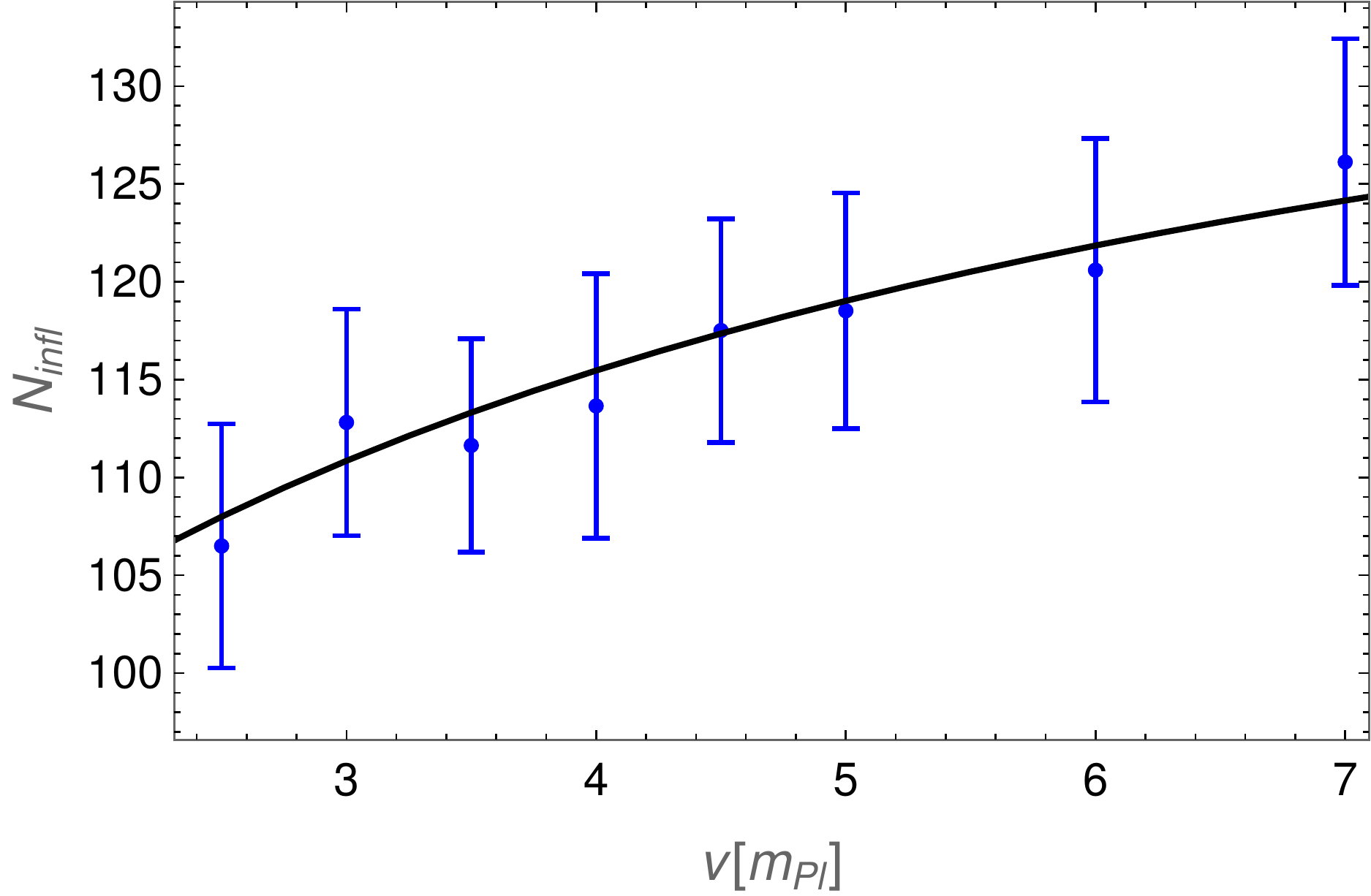}}
\subfigure[ $|\phi| < |v|$]{\includegraphics[width=7.3cm]{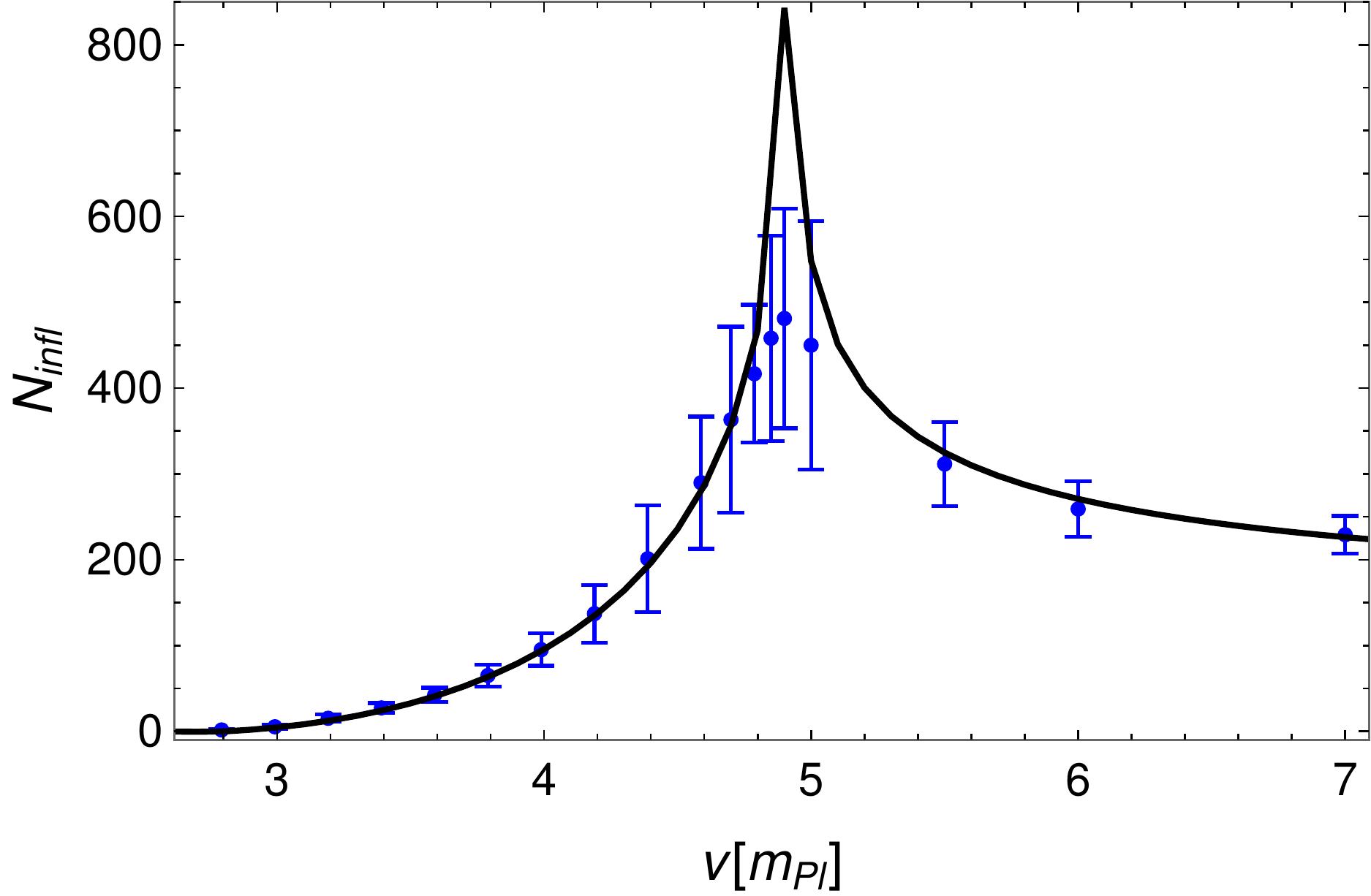} }
\subfigure[]{\includegraphics[width=7.3cm]{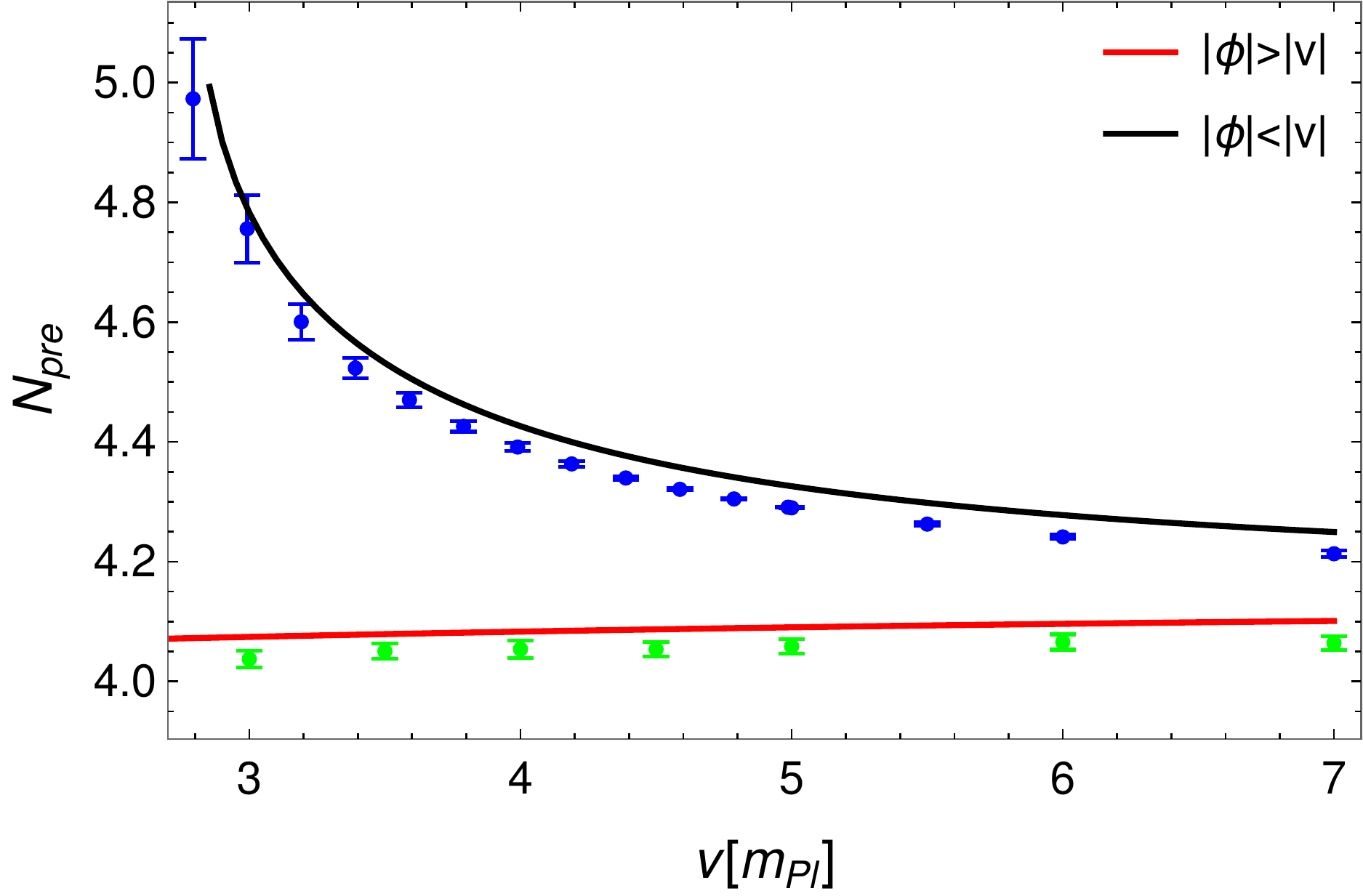} }
 \caption{The numerical results for the number of {\it e}-folds of inflation
   in the case of the Higgs-like potential when inflation happens in
   the large-field portion of the potential, $|\phi| > |v|$ [panel (a)],
   and for  inflation happening around the  plateau region,  $|\phi| <
   |v|$ [panel (b)]. The number of {\it e}-folds for the preinflationary
   regime lasting from the bounce until the beginning of inflation is
   shown in panel (c).  The data points are the results obtained
   using the methods described in Ref.~\cite{Barboza:2020jux}.  All
   curves here were obtained by setting $\bar \alpha =1/3$ within the
   procedure described in Sec.~\ref{preanalysis}.}
 \label{fig3}
 \end{figure}
\end{center}

As explained in Sec.~\ref{preanalysis}, we first need to set an
appropriate value for the ratio $\bar \alpha$ of potential to kinetic
energy in the contracting phase. We illustrate this by comparing the
results generated for the number of {\it e}-folds, Eq.~(\ref{N_folds}),
within our approach to those obtained through the statistical analysis
produced in Ref.~\cite{Barboza:2020jux}.  In
Ref.~\cite{Barboza:2020jux}, which follows the proposal initiated by
the authors of Ref.~\cite{Linsefors:2013cd},  initial conditions are
generated deep inside the contracting phase, where $\rho_\phi \ll
\rho_{\rm cr}$, and the inflaton is oscillating around the minimum of
its potential. The number of {\it e}-folds of inflation is then obtained by
taking a large number of random initial conditions satisfying these
conditions and each one of them is evolved up to the end of
inflation. The probability distribution function for each
potential is obtained, from which statistical predictions for the
number of {\it e}-folds are derived. In Ref.~\cite{Barboza:2020jux} results
were obtained for the power-law potential with $n=1,\,2,\, 3$ and also
for the Higgs-like potential for different values for the VEV $v$.
These results of Ref.~\cite{Barboza:2020jux} are indicated,  for
comparison, by the data points with the error bars shown in
{}Fig.~\ref{fig2}, for the case of the power-law potential, and in
{}Fig.~\ref{fig3}, for the case of the Higgs-like  potential.

{}From the results shown in {}Fig.~\ref{fig2}, we see that the results
very reasonably fit the data points, with a difference of less than
$5\%$, for the choice\footnote{We note that the same value for the
  constant $\alpha$ given by $1/3$ was, coincidentally, also found by
  the authors  of Ref.~\cite{Ashtekar:2011rm}, though matching their
  numerical results that were obtained in the postbounce phase.}
$\bar \alpha = 1/3$. The same choice  of $\bar \alpha$ is also seen to
reproduce well the results for the Higgs-like potential, for many
different values for $v$, as shown in {}Fig.~\ref{fig3}.  Note that
for the Higgs-like potential curves do not quite agree with the
data points for the preinflationary number of {\it e}-folds, but the
qualitative agreement is still very good, again within less than $5\%$
differences.\footnote{We notice  that for all data points shown in the
  figures, the error bars are the one-standard  deviation from the
  average values.} Hence, our results indicate that we can just fix
$\bar \alpha = 1/3$ once and for all other primordial potentials it
should also continue to produce sufficiently accurate results. This is
what we have assumed in all our subsequent results shown next.

As already said above, the results shown in {}Figs.~\ref{fig2} and
\ref{fig3}, all our subsequent analysis will be done by choosing
$\bar\alpha=1/3$. Recalling that $\alpha$ is related to the equation
of state by $w=(1-\alpha)/(1+\alpha)$, hence, the choice $\bar
\alpha=1/3$ is also equivalent  to considering the moment $t_\alpha$
in the contracting phase where $w=1/2$. 

{}Figures~\ref{fig2} and \ref{fig3} already show some revealing
features.  One recalls that  one typically requires at least around
$80$ {\it e}-folds of total expansion from the bounce to the end of
inflation in order for the quantum effects on the  spectra to be
sufficiently diluted~\cite{Zhu:2017jew}.  On the contrary, if the
total expansion lasts less than this minimum, the LQC effects on
the spectra would already be visible. This limitation, which is a
consequence of the effects of LQC on the power spectrum, will be
detailed below in the next section. {}From {}Fig.~\ref{fig2}(a),
therefore, it indicates that monomial power-law potentials with a
fifth power in the inflaton ($n=2.5$) and higher are not favored due
to the small amount of expansion predicted from them. {}For the
quadratic potential ($n=1$) the situation is quite different. We
obtain $N_{\rm infl} \sim 147$,  in agreement with  previous results
obtained in Ref.~\cite{Linsefors:2013cd}. Despite being in agreement
with current CMB data, such high value for  $N_{\rm infl}$ does
not lead to good prospects in  observing signals from the high energy
regime on CMB data. On the other hand, for the quartic model
($n=2$) we obtain  $N_{\rm infl} \sim 87$. This value, while providing a
satisfactory number of {\it e}-folds of inflation, it also allows for
better prospects concerning potentially observable signals from the
quantum regime of LQC on future CMB measurements. In 
{}Fig.~\ref{fig3}(a), which shows our results for the Higgs-like
potential when inflation happens on the large-field portion of the
potential, i.e., $|\phi| > |v|$, the number of {\it e}-folds is always
larger than $N_{\rm infl} \sim 87$, whose value is approached when $v
\to 0$ and the model is analogous to the quartic monomial
potential. However, when inflation happens in the plateau region of
the potential, i.e., when $|\phi| < |v|$, the number of {\it e}-folds tends
initially to increase with the value of the {\rm VEV} $v$, up to around $v
\sim 5 m_{\rm Pl}$, after which it drops and tends to asymptote around
$N_{\rm infl} \sim 200$. This behavior was already hinted in the
analysis done in Ref.~\cite{Barboza:2020jux} but it becomes clear now
in the results shown in {}Fig.~\ref{fig3}(b). This apparently odd
behavior has  a simple explanation. {}For small {\rm VEV} values, the
plateau region of the potential is small and it is difficult to
localize the inflaton in that small field region of the potential and
the number of {\it e}-folds of inflation tends to be small.  As the {\rm VEV}
increases, it becomes more likely for inflation to happen closer to
the flatter top of the potential and the number of {\it e}-folds of
inflation increases\footnote{It  should be noticed that the number of
  {\it e}-folds in {}Fig.~\ref{fig3}(b) appears to grow very sharp at around
  $v \simeq 5 m_{\rm Pl}$, we expect the real situation to display a
  smoother maximum as actually indicated by the numerical data. This
  is because there is always Gaussian stochastic quantum fluctuations
  acting on the background inflaton
  field~\cite{Bastero-Gil:2016mrl}. These fluctuations make unlikely
  for the inflaton to be precisely localized at the top of the maximum
  of the potential at $\phi=0$.}. However, for even larger values for
the {\rm VEV}, it becomes again less likely that the  dynamics would put the
inflaton too close to the top of the potential and the number of
{\it e}-folds decreases. Our results indicate that there is an optimum
value for the {\rm VEV} for inflation having a maximum number of {\it e}-folds for
a Higgs-like potential in the context of {\rm LQC} and this value for the
{\rm VEV} is around $v \sim 5 m_{\rm Pl}$. On the other hand, our results
also show that for $v \lesssim 3\,m_{\rm Pl}$, there are essentially
no more initial conditions leading to inflation starting and ending in
the plateau region, which also agrees with the findings of
Ref.~\cite{Bastero-Gil:2016mrl}. 

\begin{table*}[!htpb]
\centering
\caption{Numerical values obtained for the three forms of monomial
  potentials and the Starobinsky potential.}
\label{potentials table}
\begin{tabular}{cccccccccccccc}
\multicolumn{14}{c}{} \\ \hline\hline
Model &  & $V_0/m_{\rm Pl}^{4}$ &  & $\phi_{\rm B}/m_{\rm Pl}$ &  & $t_{tr}^+/t_{\rm Pl}$ &  & $\phi_{\rm tr}/m_{\rm Pl}$ &  & $(10^6) \dot{\phi}_{\rm tr}/m_{\rm Pl}^2$ & $t_{\rm i}/t_{\rm Pl}$ &  & $\phi_{\rm i}/m_{\rm Pl}$ \\ \hline
Quadratic &  & $1.355 \times 10^{-12}$ &  & $2.72$ &  & $2.9 \times 10^{4}$ &  & 4.72 &  & $5.6$ & $4.1 \times 10^{4}$ &  & 4.84 \\
Quartic &  & $1.373 \times 10^{-13}$ &  & $3.19$ &  & $2.3 \times 10^{4}$ &  & 5.22 &  & $7.1$ & $3.2 \times 10^{4}$ &  & 5.27 \\
Sextic &  & $ 4.563 \times 10^{-15}$ &  & $3.65$ &  & $2.3 \times 10^{4}$ &  & 5.68 &  & $7.1 $ & $3.1 \times 10^{4}$ &  & 5.73 \\
Starobinsky & \multicolumn{1}{l}{} & $1.497 \times 10^{-13}$ &  & $2.65$ & \multicolumn{1}{l}{} & $3.0 \times 10^{5}$ &  & 5.10 &  & $0.5$ & $4.2 \times 10^{5}$ &  & 5.16 \\ \hline \hline
\end{tabular}
\end{table*}

\begin{table*}[!htpb]
\centering
\caption{Numerical values obtained for the Higgs-like potential,
  considering some illustrative values for the VEV.}
\begin{tabular}{cccccccccccccc}
\multicolumn{14}{c}{} \\ \hline\hline
Model &  & $V_0/m_{\rm Pl}^{4}$ &  & $\phi_{\rm B}/m_{\rm Pl}$ &  & $t_{tr}^+/t_{\rm Pl}$ &  & $\phi_{\rm tr}/m_{\rm Pl}$ &  & $(10^6) \dot{\phi}_{\rm tr}/m_{\rm Pl}^2$ & $t_{\rm i}/t_{\rm Pl}$ &  & $\phi_{\rm i}/m_{\rm Pl}$ \\ \hline
\begin{tabular}[c]{@{}c@{}}Higgs-like\\ ($|\phi_{\rm B}| > |v = 3.5 m_{\rm Pl}|$)\end{tabular} &  & $ 2.867 \times 10^{-14}$ &  & $6.27$ &  & $2.4 \times 10^{4}$ &  & $8.30$ &  & $6.8$ & $3.3 \times 10^{4}$ &  & $8.35$ \\ 
\begin{tabular}[c]{@{}c@{}}Higgs-like\\ ($|\phi_{\rm B}| > |v = 4.0 m_{\rm Pl}|$)\end{tabular} &  & $ 2.384 \times 10^{-14}$ &  & $6.76$ &  & $2.4 \times 10^{4}$ &  & $8.80$ &  & $6.7$ & $3.4 \times 10^{4}$ &  & $8.85$ \\ 
\begin{tabular}[c]{@{}c@{}}Higgs-like\\ ($|\phi_{\rm B}| > |v = 4.5 m_{\rm Pl}|$)\end{tabular} &  & $ 2.010 \times 10^{-14}$ &  & $7.25$ &  & $2.5 \times 10^{4}$ &  & $9.29$ &  & $6.6$ & $3.4\times 10^{4}$ &  & $9.35$ \\ 
\begin{tabular}[c]{@{}c@{}}Higgs-like\\ ($|\phi_{\rm B}| < |v = 3.5 m_{\rm Pl}|$)\end{tabular} &  & $ 6.424 \times 10^{-14}$ &  & $-0.79$ &  & $9.0 \times 10^{4}$ &  & $1.46$ &  & $1.8$ & $1.3\times 10^{5}$ &  & $1.52$ \\
\begin{tabular}[c]{@{}c@{}}Higgs-like\\ ($|\phi_{\rm B}| < |v = 4.0 m_{\rm Pl}|$)\end{tabular} &  & $ 5.245 \times 10^{-14}$ &  & $-1.30$ &  & $6.6 \times 10^{4}$ &  & $0.90$ &  & $2.5$ & $9.4 \times 10^{4}$ &  & $0.96$ \\ 
\begin{tabular}[c]{@{}c@{}}Higgs-like\\ ($|\phi_{\rm B}| < |v = 4.5 m_{\rm Pl}|$)\end{tabular} & \multicolumn{1}{l}{} & $ 4.254 \times 10^{-14}$ &  & $-1.80$ & \multicolumn{1}{l}{} & $5.6 \times 10^{4}$ &  & $0.37$ &  & $2.9$ & $7.9 \times 10^{4}$ &  & $0.43$ \\ \hline\hline
\end{tabular}
\label{tab:my-table}
\end{table*}

Our results, including also the ones for the Starobinsky potential,
Eq.~(\ref{starobinskyV}),  are summarized in Tables~\ref{potentials
  table}-\ref{comparison_VEV}. We have chosen the cases of a
quadratic, a quartic and a sextic monomial potential for
illustration, along with some representative cases of the Higgs-like
potential and then the Starobinsky potential. 

In Tables~\ref{potentials table} and ~\ref{tab:my-table} we give the
results for the various quantities which were defined in the previous
section, in particular the numerical prediction for the amplitude for
the inflaton field at the bounce, $\phi_{\rm B}$. This is important,
since all other quantities, in particular the point where inflation
starts, depends on this value. In a sense, we see that this is
equivalent to providing the initial conditions at the bounce time. The
subsequent evolution of the Universe is then completely determined
from these initial conditions, especially the inflationary regime.  In
this case, knowing the initial conditions at the bounce, we have shown
that it completely determines the duration of inflation for any given
potential. Note that the predicted total duration of inflation here is based 
on assuming a kinetic-dominated bounce and using $\alpha=1/3$, 
differently from the statistical results obtained for example in
Refs.~\cite{Linsefors:2013cd,Linsefors:2014tna,Bolliet:2017czc,Martineau:2017sti,Barboza:2020jux}.
The predicted values for the duration of the preinflationary phase
and inflation are shown in Tables~\ref{efolds table}
and~\ref{comparison_VEV}.

\begin{table}[!htpb]
\centering
\caption{Number of {\it e}-folds obtained through the analytical
  analysis for the same models considered in Table~\ref{potentials table}.}
\label{efolds table}
\begin{tabular}{ccccc}
\hline\hline
Model &  & $N_{\rm pre}$ &  & $N_{\rm infl}$ \\ \hline
Quadratic &  & $4.15$ &  & $146.55$ \\
Quartic &  & $4.06$ &  & $86.36$ \\
Sextic &  & $4.06$ &  & $67.30$ \\
Starobinsky &  & $4.92$ &  & $1.10 \times 10^{9}$ \\ \hline\hline
\end{tabular}
\end{table}

{}From the results shown in Tables~\ref{efolds table}
and~\ref{comparison_VEV}, we can see that for all  models studied the
number of preinflationary {\it e}-folds, which consider the expansion
from the bounce to the beginning of the slow-roll inflation, is always
$N_{\rm pre} \sim 4-5$, which agrees with Ref.~\cite{Barboza:2020jux}
and other previous references. This is a consequence of the bounce being 
dominated by the kinetic energy of the  inflaton field, thus, being
weakly dependent on the form of its potential. {}For the number of
inflationary {\it e}-folds, we also obtain results generically in
agreement with those obtained in Ref.~\cite{Barboza:2020jux}. In
particular, we can observe that for the Starobinsky potential $N_{\rm
  infl} \sim 10^{9}$, i.e., the duration of slow-roll inflation is
longer compared to the other models. This is consistent with the
results shown, e.g., in Ref.~\cite{Martineau:2017sti}, which follows
the determination for the number of {\it e}-folds of inflation
originally proposed in Ref.~\cite{Linsefors:2013cd} and which was also
considered in Ref.~\cite{Barboza:2020jux}. This happens for positive
values of the scalar field at the beginning of inflation, i.e., where
the potentials have a plateau~\cite{Cicoli:2008gp}.  As discussed in
Ref.~\cite{Martineau:2017sti}, LQC dynamics automatically
provides highly energetic field configurations at the onset of
inflation. Consequently, in these cases the inflaton field is 
``pushed” away on the plateau, leading to a very long phase of 
slow-roll inflation. It is important to remember that potentials 
predicting a large number of {\it e}-folds are perfectly fine as far as
the observations are concerned.  {}For the monomial power-law potentials, as
already seen in a previous work~\cite{Barboza:2020jux}, increasing the
power $n$ implies a decreasing of the number of {\it e}-folds. 

\begin{table}[!htb]
\centering
\caption{Number of {\it e}-folds obtained for the Higgs-like  potential,
  assuming different values for the VEV.}
\begin{tabular}{ccccc}
\hline\hline Model &  & $N_{\rm pre}$ &  & $N_{\rm infl}$  \\ \hline
\begin{tabular}[c]{@{}c@{}}Higgs-like\\ ($|\phi_{\rm B}| > |v = 3.5 m_{\rm Pl}|$)\end{tabular} &  & $4.08$ &  & $113.31$ \\ \hline
\begin{tabular}[c]{@{}c@{}}Higgs-like\\ ($|\phi_{\rm B}| > |v = 4.0 m_{\rm Pl}|$)\end{tabular} &  & $4.08$ &  & $115.46$ \\ \hline
\begin{tabular}[c]{@{}c@{}}Higgs-like\\ ($|\phi_{\rm B}| > |v = 4.5 m_{\rm Pl}|$)\end{tabular} &  & $4.09$ &  & $117.35$ \\ \hline
\begin{tabular}[c]{@{}c@{}}Higgs-like\\ ($|\phi_{\rm B}| < |v = 3.5 m_{\rm Pl}|$)\end{tabular} &  & $4.53$ &  & $32.64$ \\ \hline
\begin{tabular}[c]{@{}c@{}}Higgs-like\\ ($|\phi_{\rm B}| < |v = 4.0 m_{\rm Pl}|$)\end{tabular} &  & $4.43$ &  & $95.66$ \\ \hline
\begin{tabular}[c]{@{}c@{}}Higgs-like\\ ($|\phi_{\rm B}| < |v = 4.5 m_{\rm Pl}|$)\end{tabular} &  & $4.37$ &  & $235.92$  \\ \hline\hline
\end{tabular}
\label{comparison_VEV}
\end{table}

\section{Constraining the Barbero-Immirzi parameter}
\label{additional}

As previously discussed, the Barbero-Immirzi parameter is strictly a
free parameter of the theory. Therefore, it is important to find ways
to constraint its value. In this section we study how the prediction
for the number of {\it e}-folds in LQC helps in setting possible
constraints on the Barbero-Immirzi parameter $\gamma$.

In order to constrain this parameter, we will need to analyze the power 
spectrum in LQC, as we explain in the following. In order to treat 
the perturbations in LQC we are going to adopt the so-called dressed 
metric quantization approach~\cite{Agullo:2012sh}. The dressed metric approach, 
in addition of being one of the approaches most studied in the literature, 
is the approach that seems most suitable for this kind of analysis. It  
reproduces qualitatively similar results for the power spectrum as the 
hybrid quantization approach~\cite{Fernandez-Mendez:2012poe}. In general, both 
the quantization scheme and the initial conditions chosen are important when deriving 
the power spectrum in LQC. Concerning the initial conditions, the most commonly used  
is the BD vacuum state imposed in the contracting phase, prior to the bounce. 
Since before the start of the bouncing phase all the important modes are within 
the effective horizon, the BD vacuum state is a natural choice in this case. 
A second possibility would be to impose initial conditions at the bounce. 
At the bounce some modes are inside the effective horizon and some are outside it. 
So, in this case the BD vacuum state is no longer a suitable choice. 
Instead, one can impose the fourth-order adiabatic vacuum state~\cite{Agullo:2012sh}. 
Within the validity of the latter, it has been shown in the literature~\cite{Zhu:2017jew} 
that both choices  essentially lead to the same results.

\subsection{The power spectrum in LQC}

The quantum bounce changes the scalar power spectrum by a correction
term which depends on the characteristic scale at the bounce. This
characteristic scale is the shortest scale (or largest wave number,
namely $k_{\rm B}$) that feels the spacetime curvature during the
bounce. In previous works (see, for instance,
Ref.~\cite{Benetti:2019kgw}), precise constraints on the correction
term in the scalar power spectrum of LQC was obtained using the
recent CMB data, providing limits  on the characteristic scale
$k_{\rm B}$. It happens that this scale is a function of the
Barbero-Immirzi parameter and the number of {\it e}-folds of expansion
from the bounce until today, $N_{\rm T}$. Therefore, in the following
we use such observational constraints on $k_{\rm B}$ in order to
impose limits on the value of $\gamma$ as a function of the {\it
  e}-fold number. 

In addition to the modifications at the background level from 
  LQC, at the perturbative level modifications are also expected,
especially from  relevant modes which have physical wavelengths
comparable to the curvature radius at the bounce time.  Unlike what
happens in GR, where it is usually assumed that the
preinflationary dynamics does not have any effect on modes observable
in the CMB, in LQC the situation is different. Modes that
experience curvature are excited in the Planck regime around the
bounce time. The main effect at the onset of inflation is that the
quantum state of perturbations is populated by excitations of these
modes over the BD vacuum.  As a consequence, the scalar
curvature power spectrum in LQC gets modified with respect to
GR, such that it can be written as (see Ref.~\cite{Zhu:2017jew}
for more details)
\begin{eqnarray}
\Delta_\mathcal{R}(k) &=& |\alpha_{k} +\beta_{k}|^{2}
\Delta_\mathcal{R}^{GR}(k)  \nonumber \\  &=& (1+ 2 |\beta_k|^2 + 2
      {\rm Re}(\alpha_k \beta_k^*))  \Delta_\mathcal{R}^{GR}(k).
\label{PR1}
\end{eqnarray}
In Eq.~(\ref{PR1}) $\alpha_{k}$ and $\beta_{k}$ are the Bogoliubov
coefficients, where the preinflationary effects are codified, and
$\Delta_\mathcal{R}^{GR}$ is the GR form for the power spectrum.
In GR with the BD vacuum, the Bogoliubov coefficients
in Eq.~\eqref{PR1} should reduce simply to  $\alpha_k \to
\alpha_k^{\rm BD}=1$, and $\beta_k \to\beta_k^{\rm BD} =0$.  In 
  LQC, the change of the spectrum can be seen exactly as a result of
the change of the vacuum state with respect to the GR case,
since $|\beta_k|^2 \equiv n_k$ is associated with the number of
excitations in the mode $k$.

Equation~\eqref{PR1} can also be parametrized as
\begin{equation}\label{PR}
\Delta_\mathcal{R}(k)= (1+\delta_{PL})\Delta_\mathcal{R}^{GR}(k),
\end{equation}
where the factor $\delta_{PL}$ is a scale ($k$-)dependent correction,
which, following the derivation given in Ref.~\cite{Zhu:2017jew}, is
given by
\begin{eqnarray}
\label{complete}
\delta_{PL} &=&  \left[1+\cos\left(\frac{\pi}{\sqrt{3}}\right)\right]
      {\rm csch}^{2}  \left(\frac{\pi k}{\sqrt{6} k_{B}}\right)
      \nonumber \\  &+& \sqrt{2} \sqrt{\cosh\left(\frac{2\pi
          k}{\sqrt{6}
          k_{B}}\right)+\cos\left(\frac{\pi}{\sqrt{3}}\right)}
      \cos\left(\frac{\pi}{2\sqrt{3}}\right) \nonumber \\ & \times &
               {\rm csch}^{2}\left(\frac{\pi
                 k}{\sqrt{6}k_{B}}\right)\cos(2k
               \eta_{B}+\varphi_{k}),
\end{eqnarray}
where
\begin{equation}\label{short}
\varphi_{k} \equiv \arctan \left\{\frac{{\rm
    Im}[\Gamma(a_{1})\Gamma(a_{2})\Gamma^{2}(a_{3} - a_{1} - a_{2})]}
       {{\rm Re}[\Gamma(a_{1})\Gamma(a_{2})\Gamma^{2}(a_{3} - a_{1} -
           a_{2})]}\right\},
\end{equation}
with $a_1,\,a_2,\, a_3$ defined as $a_{1,2} = (1\pm 1/\sqrt{3})/2 -i
k/(\sqrt{6} k_B)$ and $a_3=1-ik/(\sqrt{6} k_B)$.  In particular,
$\eta_B$ is the conformal time at the bounce and $k_B= \sqrt{\rho_c}
a_B \sqrt{8\pi}/m_{\rm Pl}$ is the characteristic  scale also at the
bounce.

{}From the above equations, we identify
\begin{eqnarray}
2 |\beta_k|^2 &=& \left[1+\cos\left(\frac{\pi}{\sqrt{3}}\right)\right]
{\rm csch}^{2}  \left(\frac{\pi k}{\sqrt{6} k_{B}}\right), \\ 2 {\rm
  Re}(\alpha_k \beta_k^*) &=&  \sqrt{2} \sqrt{\cosh\left(\frac{2\pi
    k}{\sqrt{6} k_{B}}\right)+\cos\left(\frac{\pi}{\sqrt{3}}\right)}
\nonumber \\ & \times & \cos\left(\frac{\pi}{2\sqrt{3}}\right) {\rm
  csch}^{2}\left(\frac{\pi k}{\sqrt{6}k_{B}}\right)\cos(2k
\eta_{B}+\varphi_{k}).  \nonumber \\
               \label{complete2}
\end{eqnarray}
The term $\cos(2k \eta_{B}+\varphi_{k})$ in Eq.~(\ref{complete2})
oscillates very fast, having a negligible effect when averaging out in
time.  Therefore, for  practical purposes, in observable quantities
the factor $\delta_{PL}$ can be simply considered as being given by
\begin{equation} \label{delta}
\delta_{PL} = \left[1+\cos\left(\frac{\pi}{\sqrt{3}}\right)\right]
      {\rm csch}^{2}\left(\frac{\pi k}{\sqrt{6}k_{B}}\right).
\end{equation}
Note that in this case $\delta_{\rm PL}$ can simply be identified with
$2n_k$, i.e., with the number of excitations in the mode $k$ which
appears as a consequence of the quantum bounce in LQC.  It is
due to this correction term  that in LQC, in order for the
spectrum to be consistent with the observations, it is required that
the Universe must have expanded some extra $21$ {\it e}-folds such as
to allow  for those scale-dependent features in the primordial scalar
power spectrum to get sufficiently diluted and as discussed in detail
in Refs.~\cite{Zhu:2017jew,Benetti:2019kgw}.

Even though the correction given by Eq.~(\ref{delta}) was derived in the 
dressed metric quantization approach, the result is also qualitatively
similar when derived in the hybrid quantization approach
(see, e.g., Ref.~\cite{Li:2020mfi}). 
Other alternative quantization schemes used in
{\rm LQC} can lead to corrections to the power spectrum that are
suppressed. This seems to be the case, for example, in the
closed/deformed algebra approach~\cite{Li:2018vzr,SVicente:2022ebm}.
In this case there is no need in principle of the additional extra
{\it e}-folds of expansion as required in the dressed or hybrid
quantization approaches. Either way, we can view our results using
the dressed metric approach as the one giving the most restrictive
condition on the required minimum number of {\it e}-folds.  Even in other
approaches that might lead to a suppressed correction to the power
spectrum, one still needs to require  that inflation lasts around
some $60$ or so {\it e}-folds. This still gives a restriction qualitatively
similar to what we will consider below, though weaker.

\subsection{The Barbero-Immirzi parameter as a function of the
  number of {\it e}-folds}

The total number of {\it e}-folds of expansion from the moment of the
bounce until today, $N_{\rm T}$, is related to the LQC parameter
$k_{\rm B}$.  By assuming an upper bound on $k_{\rm B}$, it can be
translated into constraints on the total number of {\it e}-folds. We
are interested in finding an upper bound value for $k_{\rm B}$, which
implies in a lower value for the number of {\it e}-folds.  In
Ref.~\cite{Benetti:2019kgw}, for example, constraints on the parameter
$k_{\rm B}$ were obtained from CMB data. Since $k_{\rm B}$ is
related to $\gamma$, these results can be translated into constraints
in the parameter $\gamma$. 

The relation between $k_{B}$ and the number of {\it e}-folds is given
by the equation
\begin{equation}\label{kbN}
k_{\rm B}\equiv \frac{\sqrt{8\pi \rho_{\rm cr}} a_{\rm B}}{m_{\rm
    Pl}}= m_{\rm Pl}\left(\frac{\sqrt{3}}{4 \pi
  \gamma^{3}}\right)^{1/2} e^{-N_{\rm T}},
\end{equation}
where we have used Eq.~(\ref{rhoc}) and in the above equation  $N_{\rm
  T} = \ln(a_0/a_{\rm B})$ is the total number of {\it e}-folds from
the bounce until today. Note that  in Eq.~(\ref{kbN}) we have used the
standard convention of setting the  scale factor today as one,
$a_0=1$.

It happens that the CMB observations constrain directly the
value of $k_{B}$ by imposing a limit in the correction term given by
Eq.~(\ref{delta}).  An updated observation constraint on $k_{B}$ was
obtained in Ref.~\cite{Benetti:2019kgw}, which leads to $k_{B} < 1.9
\times 10^{-4} \ {\rm Mpc}^{-1}$ at $1 \sigma$. Note that this
constraint is independent of the value of the Barbero-Immirzi
parameter and should be valid for any value for $\gamma$.  Using
Eq.~(\ref{kbN}), it translates into a lower limit on $N_{\rm T}$
depending on the value for the Barbero-Immirzi parameter, which is
described by
\begin{equation}\label{22}
 N_{\rm T} \gtrsim 139 - \frac{3}{2} \ln (\gamma).    
\end{equation}
Note that $N_{\rm T}$ can be expressed as
\begin{eqnarray}
N_{\rm T} &=& \ln \left(\frac{a_0}{a_{\rm B}} \right) = \ln \left(
\frac{a_{\rm i}}{a_B} \frac{a_{\rm end}}{a_{\rm i}} \frac{a_{\rm
    reh}}{a_{\rm end}} \frac{a_0}{a_{\rm reh}} \right) 
\nonumber \\ 
&=& N_{\rm pre} + N_{\rm infl} + N_{\rm reh} + \ln
\left(\frac{a_0}{a_{\rm reh}} \right),
\label{NT}
\end{eqnarray}
where $a_i$, $a_{\rm end}$ and $a_{\rm reh}$ are the scale factors at
the beginning of inflation, at the end of inflation and at the end of
the reheating phase, respectively, while $N_{\rm reh}$ is the duration
of the reheating phase. We also have that~\cite{Munoz:2014eqa}
\begin{equation}
\frac{a_0}{a_{\rm reh}} = \left(\frac{11 g_{s,\rm reh}}{43}
\right)^{\frac{1}{3}} \frac{T_{\rm reh}}{T_{\rm CMB,0}},
\label{a0areh}
\end{equation}
where $T_{\rm reh}$ is the reheating temperature, $T_{\rm CMB,0}$ is
the temperature of the CMB today and $g_{s,\rm reh}$ is the
effective number of relativistic degrees of freedom for entropy at the
end of reheating. Considering the case of instantaneous reheating at
the end of inflation (i.e., neglecting the typically unknown physics
at reheating),  $N_{\rm reh} \approx 0$ and we can associate $T_{\rm
  reh}$  with the inflaton potential energy density at the end of
inflation, $V_{\rm end}$, as
\begin{equation}
T_{\rm reh} \simeq \left( \frac{30}{g_{\rm reh} \pi^2}
\right)^{\frac{1}{4}} (1+\kappa)^{\frac{1}{4}}  V_{\rm
  end}^{\frac{1}{4}},
\label{Treh}
\end{equation}
where $g_{\rm reh}$ is the effective number of relativistic degrees of
freedom for energy at full thermalization. In Eq.~(\ref{Treh}),
$\kappa$ is the ratio of kinetic energy to potential energy during
inflation. At the end of inflation, $\kappa=1/2$. By taking both
$g_{s,\rm reh}$ and $g_{\rm reh}$ to be close to those for the
standard model of particle physics, $g_{s,\rm reh} \simeq g_{\rm reh}
\sim 100$, we obtain that $\ln \left(a_0/a_{\rm reh}\right) \sim
60$. Thus,  Eq.~(\ref{22}) can also be written as a lower bound for
$N_{\rm pre} + N_{\rm infl}$,
\begin{equation}\label{Nbound}
N_{\rm pre} + N_{\rm infl} \gtrsim 79 - \frac{3}{2} \ln (\gamma).    
\end{equation}
Adding the reheating details after inflation only makes the above
relation more restrictive.  Thus, we can use  Eq.~(\ref{Nbound}) as an
overall lower bound for the total number of {\it e}-folds from the
bounce until the end of inflation as a function of the Barbero-Immirzi
parameter.  By considering the value for $\gamma$ as given by the
value suggested by the black hole entropy~\cite{Meissner:2004ju},
$\gamma \simeq 0.2375$, we then obtain that $N_T \gtrsim 141$ and
$N_{\rm pre} + N_{\rm infl} \gtrsim 81$.  Recalling also that the
number of {\it e}-folds relevant from the {\rm CMB} observations (e.g., at a pivot
scale $k_*=0.05/{\rm Mpc}$ and assuming instant reheating for
simplicity)  is given by~\cite{Planck:2018jri}
\begin{equation}
N_* \approx 57 + 2\ln\left(\frac{V_*^{1/4}}{10^{16}{\rm GeV}}\right) -
\ln\left(\frac{T_{\rm reh}}{10^{16}{\rm GeV}}\right) ,
\label{N*}
\end{equation}
which typically leads to $N_* \sim 50-60$ for the necessary number
of {\it e}-folds of inflation.  As we have seen that $N_{\rm pre} \sim 4$,
which is very weakly dependent on the form of the inflaton potential,
then $N_*$ can comfortably fit in the estimated lower bound $N_{\rm
  pre} + N_{\rm infl} \gtrsim 81$.

In the following we analyze the general behavior of the number of
{\it e}-folds $N_{\rm pre+infl}=N_{\rm pre} + N_{\rm infl}$ lasting
from the bounce until the end of inflation, as a function of
$\gamma$. The general expressions for  $N_{\rm pre}$ and $N_{\rm
  infl}$ were derived in the previous section. Next, we will consider
how $N_{\rm pre+infl}$ changes by varying the Barbero-Immirzi
parameter and also by considering the overall lower bound  given by
Eq.~(\ref{Nbound}).

\subsection{Results for the Barbero-Immirzi parameter}

\begin{figure}[!htb]
\centering \subfigure[]{\includegraphics[width=7.5cm] {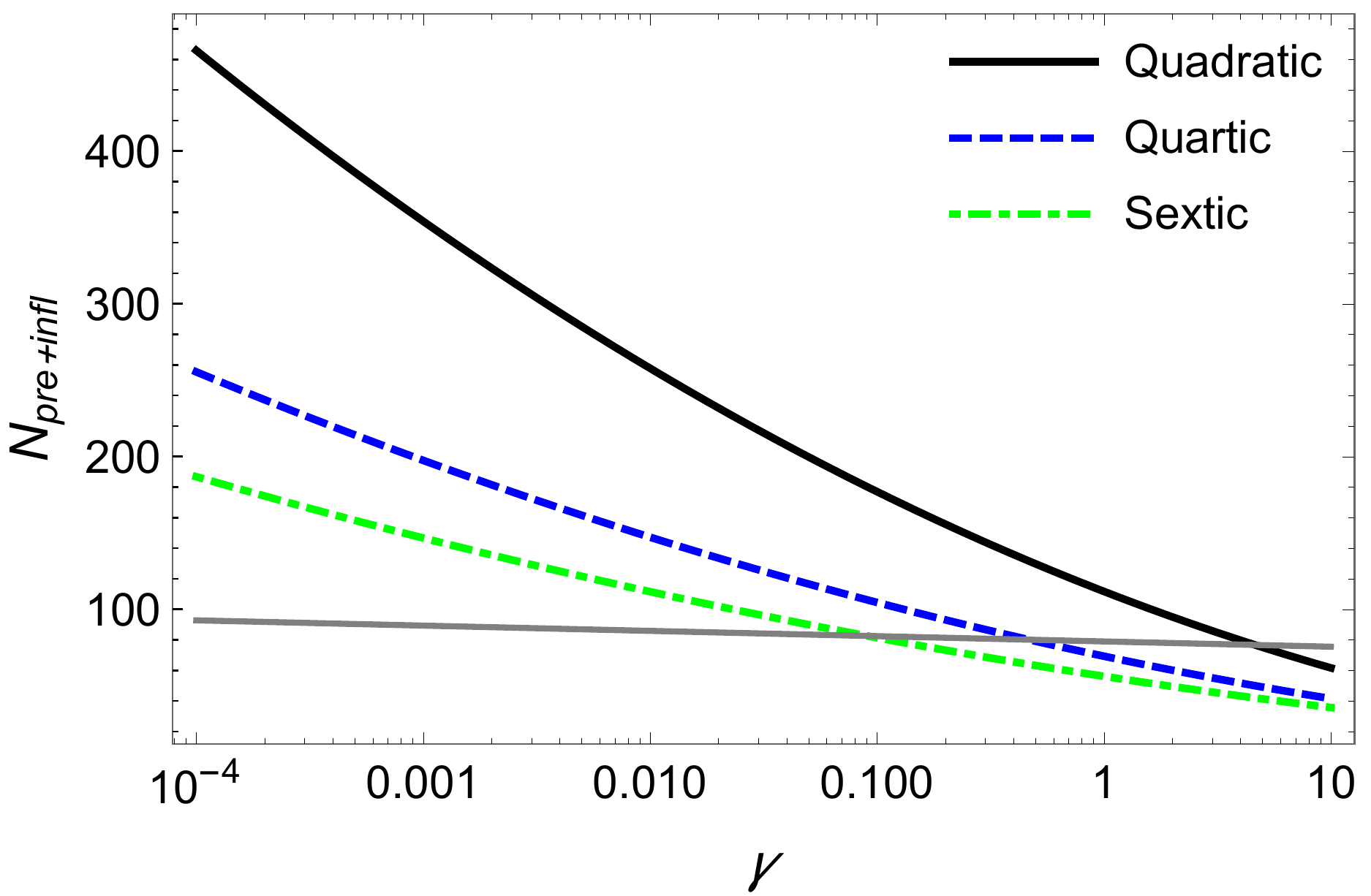}}
\subfigure[]{\includegraphics[width=7.5cm] {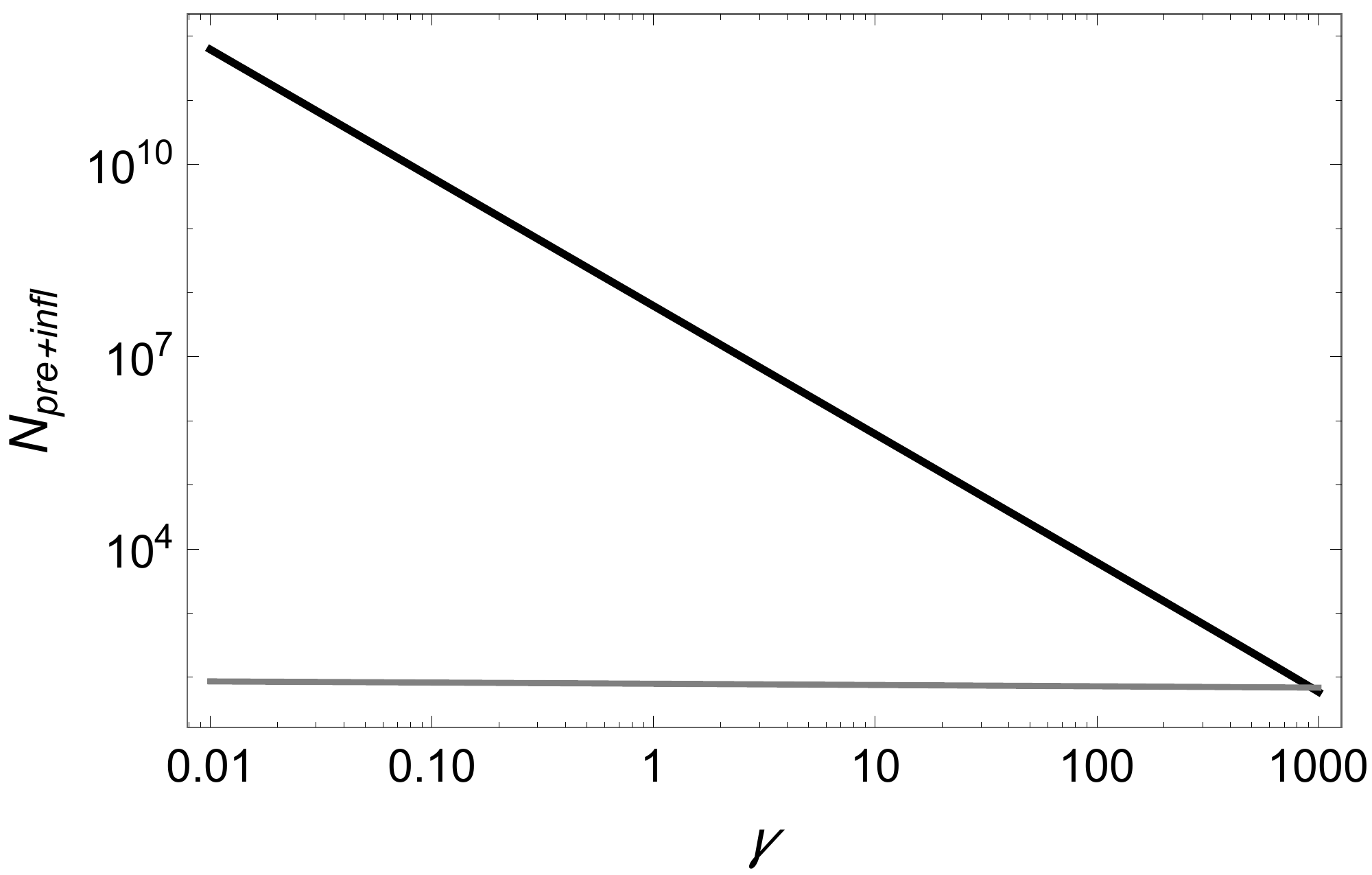}}
\caption{Number of preinflationary plus  inflationary {\it e}-folds,
  $N_{\rm pre+infl}$,  for the  potentials considered, varying the
  Barbero-Immirzi parameter. The almost horizontal gray lines  show
  the lower limit on $N_{\rm  pre+infl}(\gamma)$ set by
  Eq.~(\ref{Nbound}). Panel (a): power-law potentials; panel(b):
  Starobinsky potential.}
\label{POTENTIALS_gamma}
\end{figure}

\begin{figure}[!htb]
\centering \subfigure[]{\includegraphics[width=7.5cm] {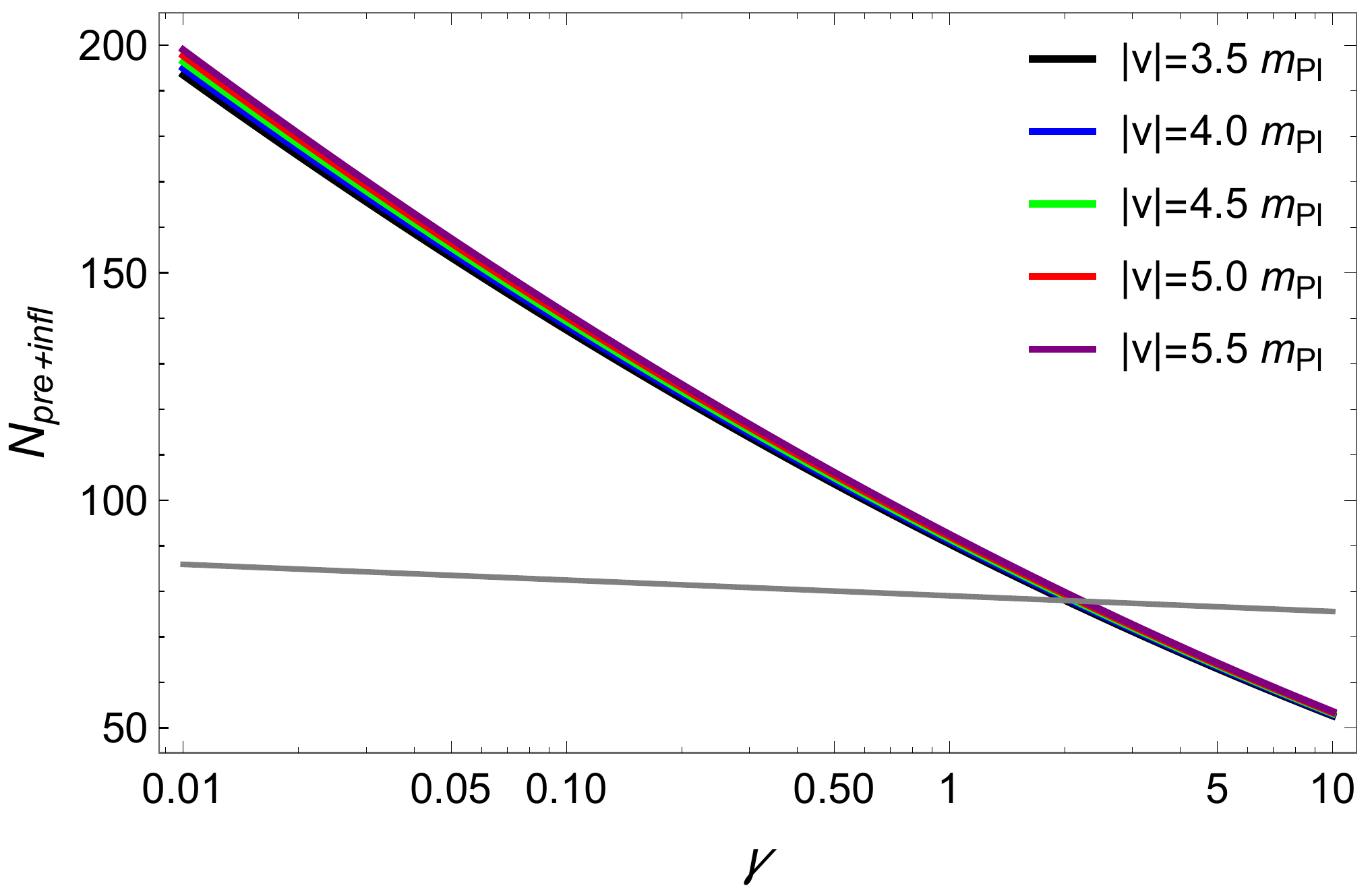}}
\subfigure[]{\includegraphics[width=7.5cm] {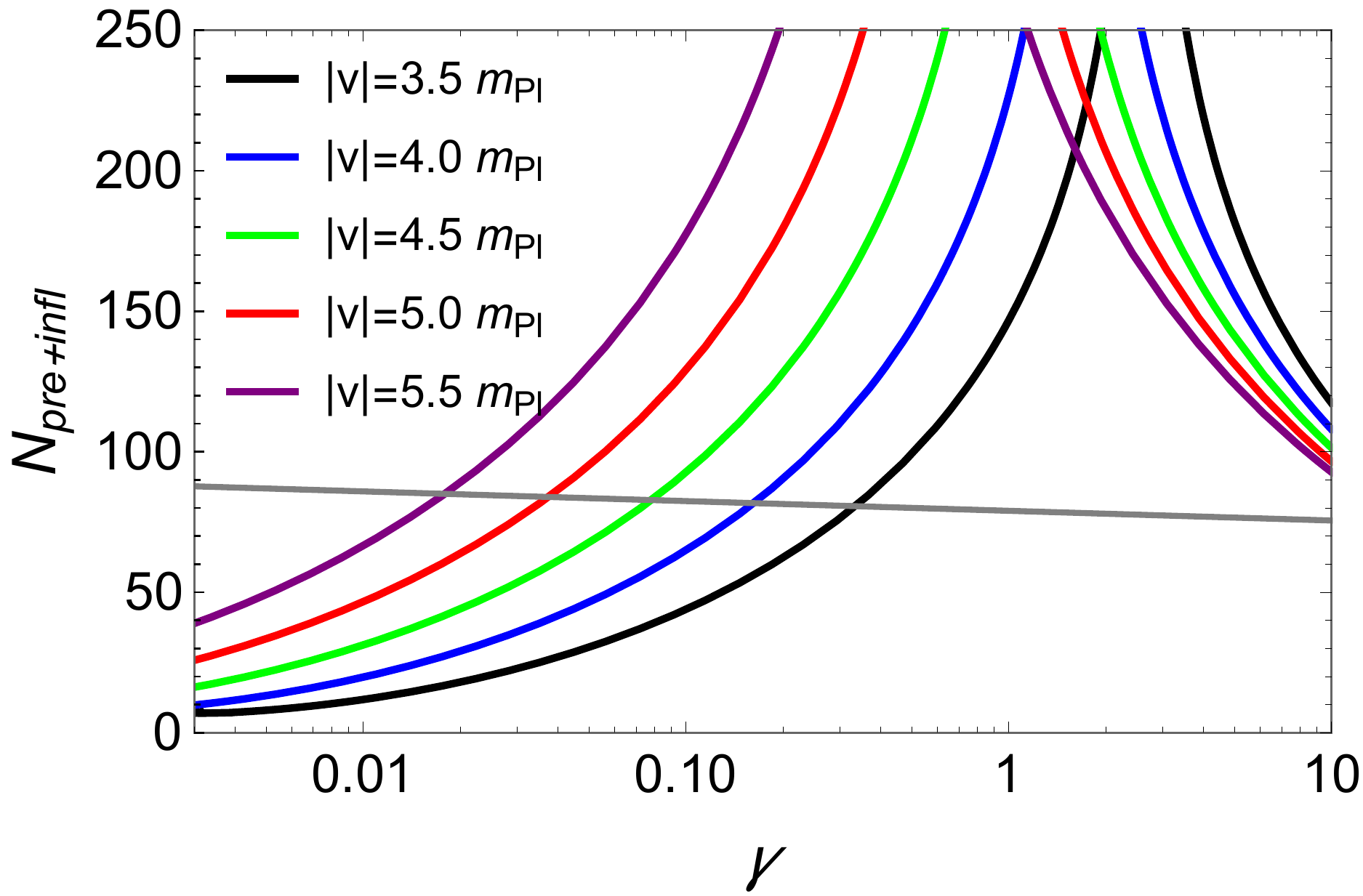}}
\caption{Number of  {\it e}-folds, $N_{\rm  pre+infl}$, as a function
  of  the Barbero-Immirzi parameter for the Higgs-like  potential with
  different values of the VEV. The almost horizontal gray lines
  show the lower limit on $N_{\rm  pre+infl}(\gamma)$ set by
  Eq.~(\ref{Nbound}). Panel (a): the results for the large field case
  $|\phi| > |v|$. Panel (b): the results for the small field case
  $|\phi|< |v|$.}
\label{NinflChaotic_Ninflplateau}
\end{figure}

The behavior of the number of {\it e}-folds from the bounce to the end
of inflation as a function of  $\gamma$ for each  potential considered
in this paper is shown in {}Figs.~\ref{POTENTIALS_gamma} and
\ref{NinflChaotic_Ninflplateau}.  We show in
    {}Fig.~\ref{POTENTIALS_gamma}(a) the case of monomial power-law
    potentials.  We consider a sufficiently long range of
    representative values for $\gamma$ (the range $0 \lesssim \gamma
    \lesssim 10$ typically corresponds to the interval most considered
    in the
    literature~\cite{Asante:2020qpa,Asante:2021zzh,Perlov:2020cgx,Broda:2010dr,Mercuri:2009zt}). The
    almost horizontal gray  lines  in {}Figs.~\ref{POTENTIALS_gamma}
    and \ref{NinflChaotic_Ninflplateau} show the lower limit on
    $N_{\rm  pre+infl}(\gamma)$ set by Eq.~(\ref{Nbound}). In all
    cases  it is observed that the smaller the value of $\gamma$,  the
    greater the value of $N_{\rm pre+infl}$.
{}For the quadratic case $(n=1)$ we obtain a value of $N_{\rm  pre+infl}$ which satisfies the
observational constraints for
    almost the entire interval of $\gamma$. However, the quartic
    $(n=2)$ and sextic $(n=3)$ cases approach the limit of $81$ {\it
      e}-folds for smaller values of $\gamma$. {}For the quartic case
    $(n=2)$, we get that for  $\gamma \sim 0.46$, $N_{\rm \rm
      pre+infl}$ reaches the value of $81$ {\it e}-folds. {}For the
    sextic case~$(n=3)$, we get that for  $\gamma \sim 0.1$, the
    number of {\it e}-folds reaches the limiting value $N_{\rm
      pre+infl}= 81$ and quickly drops below this value as $\gamma$
    increases. We can see that, although for the usual value of the
    Barbero-Immirzi parameter the sextic potential in LQC is in
    strong tension with the observations, for smaller values of the
    parameter, $\gamma < 0.1$, it can be consistent.

Complementing these results, we show in
{}Fig.~\ref{POTENTIALS_gamma}(b) the results for the number of {\it
  e}-folds as a function of $\gamma$ for the Starobinsky potential. We
can see that, again, this corresponds to the case that presents the
highest values for $N_{\rm  pre+infl}$. \footnote{See, for example,
  Ref.~\cite{Martineau:2017sti} for an analysis on the duration of the
  slow-roll inflation for this kind of potential in the presence of
  shear.} We analyzed what value for the $\gamma$ parameter would
lead to the limiting value of $N_{\rm  pre+infl} = 81$. The numerical
results  shows that in this case the value of the Barbero-Immirzi
parameter would be around $\gamma \sim 1000$, as can be seen in
{}Fig.~\ref{POTENTIALS_gamma}(b). Note that  although $\gamma$ is typically expected to be approximately of order $1$, a large $\gamma$ is not inconsistent.

To complete our analysis, the results for the Higgs-like  potential is
shown in {}Fig.~\ref{NinflChaotic_Ninflplateau}. The number of {\it
  e}-folds as a function of $\gamma$  for  the large field case is
shown in {}Fig.~\ref{NinflChaotic_Ninflplateau}(a), while for the
small field case it is shown in
{}Fig.~\ref{NinflChaotic_Ninflplateau}(b). In both cases,
representative values for the {\rm VEV} are considered. We observe
from  {}Fig.~\ref{NinflChaotic_Ninflplateau}(a), which is for the
large field case, that $N_{\rm  pre+infl}$ decreases when  $\gamma$
increases. {}For all {\rm VEVs} considered, we observe a similar behavior
and we see that the limit  $N_{\rm  pre+infl}=81$  occurs when $\gamma
\approx 1.15$. On the other hand, for the small field case, shown in
        {}Fig.~\ref{NinflChaotic_Ninflplateau}(b), it is possible to
        observe that the higher the {\rm VEV}, the higher  the
        number of {\it e}-folds  for each value of $\gamma$, up to
        some critical value for $\gamma$, above which the number of
        {\it e}-folds decreases. The decreasing behavior for the number of
        {\it e}-folds seen here, and that happens above some value of
        $\gamma$,  is a consequence of the effect already discussed in
        connection to the one seen in {}Fig.~\ref{fig3}(b).
        Increasing $\gamma$ changes the point ({\rm VEV}) where the number
        of {\it e}-folds peaks when inflation happens in the plateau region
        of the potential. {}For {\rm VEVs} larger than around $7 m_{\rm
          Pl}$, the number of {\it e}-folds drops below the lower bound set
        by Eq.~(\ref{Nbound}) for $\gamma \sim 10$, but for even
        higher {\rm VEVs}, the lower bound will be reached for much smaller
        values of $\gamma$.

The results given above show that for different values of $\gamma$ the
predictions for observational signals of LQC in CMB  are
considerably different. We see that the consistency of the models with
the current data depends strongly on the value of this parameter. This
motivates and highlights the importance of a careful analysis of the
role of the Barbero-Immirzi parameter in these scenarios.

\section{Conclusions}
\label{conclusion}

It is well known that the primordial power spectrum is a quantity that
can relate the theory of the early Universe with the
observations. Previous works in the literature have shown that the
predictions for the power spectrum in the context of LQC receive
a correction term with respect to the predictions in the context of
GR. As shown in  Ref.~\cite{Zhu:2017jew}, due to this correction
term, models in the context of LQC require a minimum amount of
$\sim 81$ {\it e}-folds of expansion from the bounce until the end of
the slow-roll inflationary phase in order to be consistent with the
observations. However, the prediction for the duration of inflation in
LQC depends on how the initial conditions for the dynamics are
set. 

A previous work~\cite{Barboza:2020jux} investigated in detail the duration of inflation for LQC models with monomial
and Higgs-like  potentials and which was considered initial conditions
far in the contracting phase, thus well before the bounce (extending
the analysis made in
Refs.~\cite{Ashtekar:2011rm,Zhu:2017jew,Chen:2015yua,Linsefors:2013cd,Linsefors:2014tna,Bolliet:2017czc,Martineau:2017sti,Shahalam:2017wba,Shahalam:2019mpw,Sharma:2018vnv}).
In the present paper, we have investigated the duration of inflation
in these same models, including also the Starobinsky potential, but
now from a different perspective concerning the initial conditions. As
discussed, the dynamics in the LQC models considered here starts in
the contracting phase sufficiently before the bounce, such that the
kinetic energy of the inflaton field necessarily comes to dominate at
the bounce. Then, we have shown that it is possible to estimate the
inflaton field amplitude at some intermediate instant in the
contracting phase, but still well before the bounce. With that value
for the inflaton amplitude, we can forward the background dynamics up
to the bounce instant and determine the value for the inflaton field
at that instant, $\phi_{B}$. With $\phi_{B}$ uniquely determined, all
subsequent background dynamics from the bounce until the end of
inflation can then be determined. In LQC models in which the evolution
of the inflaton field is dominated by its kinetic energy at the
quantum bounce, a slow-roll inflation phase is practically always
reached as is also demonstrated by our results.
With the initial conditions taken deeper in the contracting phase, the kinetic 
energy regime dominates earlier and longer until the bounce is reached. 
Hence, we expect the results to turn out to be weakly dependent on the 
inflationary potential, as our results indicate. This also implies that 
the results turn out to be weakly dependent on the specific value 
adopted for the ratio of the kinetic and potential energy.

{}For all the models analyzed, we found that the number of {\it
  e}-folds of the preinflationary phase is approximately $N_{\rm pre}
\sim 4$. On the other hand, the number of inflationary {\it e}-folds
changes considerably depending on the potential for the
inflaton. Monomial potentials like $V \propto |\phi|^5$ and with
higher powers tend to predict a too small value for the number of
inflationary {\it e}-folds and, thus, they are likely to be
incompatible with the CMB data. The quartic potential, $V\propto
\phi^4$, on the other hand, predicts the most likely $N_{\rm infl}$ to
be around $N_{\rm infl} \sim 86$, which suggests a very good
possibility of leading to observable signatures from LQC in the
spectrum of CMB data. {}For the quadratic model, $V\propto
\phi^2$, the most likely $N_{\rm infl}$ is around $N_{\rm infl} \sim
147$. This is in agreement with the results obtained in the earlier
work done in Ref.~\cite{Linsefors:2013cd}.  With such high values of
$N_{\rm infl}$ allowed by the quadratic potential, the effects from
the quantum regime would probably be diluted to an unobservable
level. {}For the Higgs-like symmetry-breaking potential we have shown
that $N_{\rm infl}$ grows with the VEV $v$ for the case of inflation
occurring in the plateau~(small-field) region. It reaches a maximum
value for the number of {\it e}-folds at a VEV around $v\sim 5 m_{\rm Pl}$
and beyond this value $N_{\rm infl}$ drops and tends to asymptote at
around $N_{\rm infl} \sim 200$. {}For inflation occurring in the
large-field ($|\phi| > |v|$) part of the potential $N_{\rm infl}$  has
a weak dependence on $v$, being around $N_{\rm infl} \sim 110$ in the
range of values of $v$ we have considered. Even though these results
were obtained for a Higgs-like potential, we expect the features
displayed would also be present in other small field types of
potentials, like hilltop and axionlike potentials. {}For the
Starobinsky model, we predict a much higher value for $N_{\rm infl}$
when compared to the other potentials studied, with $N_{\rm infl} \sim
10^{9}$. This implies no potentially observable signal that could
be searched for on CMB data as far as the Starobinsky model is
considered in the context of LQC. 

We have also shown in the present paper that the value of the
Barbero-Immirzi parameter can affect strongly the constraints on the
number of {\it e}-folds. The Barbero-Immirzi parameter is strictly a
free parameter of the underlying LQG theory. Being related to
the typical scale at the bounce, the Barbero-Immirzi parameter implies
in different predictions for the power spectrum in LQC. In fact,
what CMB actually constrains is the  combination of the
parameters $\gamma$ and $N_{\rm pre+infl}$, the total number of {\it e}-folds
from the bounce until the end of inflation. Therefore, it is
important to investigate the relation between the predictions for the
number of {\it e}-folds in LQC with the value of $\gamma$. This
analysis is performed in detail, for the first time, in the present
paper.

The results presented in this paper show that for different values of
$\gamma$ the predictions for the duration of inflation in LQC
are considerably different. {}For the monomial potentials, the
predicted number of {\it e}-folds decreases with the value of
$\gamma$. In particular it is interesting to see that, for example in
the case of the sextic potential in LQC, although for the usual
value of the Barbero-Immirzi parameter the model is in strong tension
with the observations, for smaller values of the parameter, like
$\gamma\lesssim 0.1$, it can be consistent with the increase of the
predicted number of {\it e}-folds when $\gamma$ is lower than the usual
value adopted for it in the literature. {}For the Higgs-like
potential, we have obtained that the number of {\it e}-folds increases
with $\gamma$ in the small field case, up to some critical value,
beyond which, with the increase of $\gamma$, it starts to
decrease. {}For the large field case, on the other hand, the
number of {\it e}-folds always  decreases when $\gamma$ increases.
{}For the Starobinsky model, we have obtained that the prediction for 
the number of {\it e}-folds decreases with $\gamma$. The number of {\it e}-folds  
can reach the limiting value of $N \simeq 81$ for the value of the Barbero-Immirzi 
parameter $\gamma \sim 1000$, which is, nevertheless, a quite high value to 
be acceptable by the underlying LQG theory.

It is important to remark that  the observable predictions in LQC 
and the constraints obtained for the Barbero-Immirzi parameter are dependent on the 
way initial conditions are set. In this work we considered the initial conditions 
for the perturbations to be the BD vacuum in the contracting phase in the context of the 
dressed metric approach,  which leads to basically the same results as considering the 
fourth-order adiabatic vacuum state at the bounce in the context of the dressed metric approach.
We  obtain that the observable predictions in LQC models are dependent 
also on the value of the Barbero-Immirzi parameter, 
being quite sensitive to the latter. We obtain limits  for $\gamma$ directly from the 
primordial scalar power spectrum. Since the consistency of the models with the current data
depends strongly on the value of this parameter, this paper highlights
for the first time the importance of a careful analysis of the role of
the Barbero-Immirzi parameter in LQC.

Finally, in this paper
we have analyzed only the case of a Universe with energy density made
essentially of the inflaton field. As possible follow-ups of this
paper, it would be interesting to perform a similar analysis when
other energy contents are present, like from sources of
anisotropies~\cite{Linsefors:2014tna,Martineau:2017sti}  or when
radiation is also present, which itself has been shown to lead to
interesting results in the context of
LQC~\cite{Graef:2018ulg,Barboza:2020jux,Benetti:2019kgw,Herrera:2010yg,Xiao:2011mv,Zhang:2013yr,Xiao:2020aij}.

\section{ACKNOWLEDGMENTS}

 L.N.B. acknowledges financial support of the Coordena\c{c}\~ao de
 Aperfei\c{c}oamento de Pessoal de N\'{\i}vel Superior (CAPES) -
 Finance Code No. 001. G.L.L.W.L. acknowledges financial support of the
 Conselho Nacional de Desenvolvimento Cient\'{\i}fico e Tecnol\'ogico
 (CNPq).   L.L.G is supported by CNPq, under the Grant
 No. 307052/2019-2, and by the Funda\c{c}\~ao Carlos Chagas Filho de
 Amparo \`a Pesquisa do Estado do Rio de Janeiro (FAPERJ), Grant
 No. E-26/201.297/2021. R.O.R. is partially supported by research
 grants from CNPq, Grant No. 307286/2021-5, and from FAPERJ, Grant
 No. E-26/201.150/2021. R.O.R. also acknowledges financial support of
 the  Coordena\c{c}\~ao de  Aperfei\c{c}oamento de Pessoal de
 N\'{\i}vel Superior (CAPES) -  Finance Code No. 001.  One of us (L.L.G.)
 wishes to thank the  Kavli Institute for the Physics and Mathematics
 of the Universe (IPMU) for kind hospitality.
 
\appendix

\section{Obtaining $V_{0}$ through the CMB spectrum}  
\label{appendix_V0}

Let us now briefly review the derivation of the normalization $V_0$
for each of the potentials we have considered in this paper.  The
primordial scalar curvature power spectrum $\Delta_{{\cal R}}$ is
given by the standard expression~\cite{Lyth:2009zz}
\begin{eqnarray} \label{PkCI}
\Delta_{{\cal R}} &=&  \left(\frac{ H_*^2}{2
  \pi\dot{\phi}_*}\right)^2,
\end{eqnarray}
where a subindex $*$ means that the quantities are evaluated at the
Hubble radius crossing  $k_*$  ($k_*=a_* H_*$).  This is typically
assumed to happen around $N_* \sim 50-60$ {\it e}-folds before the
end of inflation. In this work we have assumed the fiducial value of
$60$ {\it e}-folds for illustration purposes. The value of $V_{0}$ is fixed by
the normalization of the primordial scalar of curvature power
spectrum. The Planck collaboration~\cite{Aghanim:2018eyx} gives for
instance the value $\ln\left(10^{10} \Delta_{{\cal R}} \right) \simeq
3.047$ (TT,TE,EE-lowE+lensing+BAO 68$\%$ limits). This is the value we
have adopted in this paper to obtain the normalization $V_0$.

During the slow-roll regime of inflation, we can make the
approximations  $H^{2} \simeq 8\,\pi V/( 3\,m_{\rm Pl}^{2})$ and
$\dot{\phi} \simeq -V_{,\phi}/(3\,H)$. Thus,
\begin{equation}\label{spectrum_amplitude}
 \Delta_{\rm R} \simeq \frac{128\,\pi}{3\,m_{\rm Pl}^{6}}
 \frac{V_{\ast}^{3}}{V_{,\phi \ast}^{2}} , 
\end{equation}
for any given potential. 

{}For the monomial power-law potentials [Eq.~\eqref{powerlawV}],
Eq.~(\ref{spectrum_amplitude}) gives 
\begin{equation}\label{pwl_spectrum}
 \Delta_{\rm R} = \frac{4}{3\,(4 \pi)^{n}} \frac{1}{n^{3}}
 \frac{V_{0}^{\rm mon}}{m_{\rm Pl}^{4}}\left[n (2\,N_{\ast} +
   n)\right]^{n+1},
\end{equation}
where we have used that 
\begin{equation}
\phi_{\ast}\equiv \phi(N_*)=\sqrt{(n\,m_{\rm Pl}^{2}/4\,\pi)
  \left(2\,N_{\ast} + n\right)}.
\end{equation}  
Therefore, the normalization $V_{0}$ obtained from CMB
measurements gives
\begin{equation}
   \frac{V_{0}^{\rm mon}}{m_{\rm Pl}^{4}} = \frac{3\,(4\,\pi)^{n}}{4}
   \frac{n^{3}}{\left[n (2\,N_{\ast} + n)\right]^{n + 1}}\,
   \Delta_{\rm R}. 
\end{equation}

{}For the Higgs-like potential~\eqref{higgsV}, we find, analogously,
that
\begin{equation}\label{Delta_R_CMB}
    \Delta_{\rm R} = \frac{2\,\pi}{3\,m_{\rm Pl}^{6}} \frac{V_{0}^{\rm
        Higgs}}{m_{\rm Pl}^{4}} \frac{(\phi_{\ast}^{2} -
      v^{2})^{4}}{\phi_{\ast}^{2}}.
\end{equation}
Solving for $V_{0}$, we find
\begin{equation}
 \frac{V_{0}^{\rm Higgs}}{m_{\rm Pl}^{4}} = \frac{3\,m_{\rm
     Pl}^{6}}{2\,\pi}\,\Delta_{\rm
   R}\,\frac{\phi^{2}(N_{\ast})}{\left[\phi_i^{2}(N_{\ast}) -
     v^{2}\right]^{4}},
\end{equation}
where $\phi(N_{\ast})$ is obtained from the
expression~\eqref{ninf_higgs}, which gives two possible solutions,
\begin{equation}
 \phi^{2}(N_{\ast}) = - v^{2}\,W_{0} \left[-
   \frac{\phi_{end}^{2}}{v^{2}} \left(e^{\frac{N_{\ast}}{\pi} +
     \frac{\phi_{\rm end}^{2}}{m_{\rm Pl}^{2}}}\right)^{-\frac{m_{\rm
         Pl}^{2}}{v^{2}}}\right]  
\label{V0higgs1}
\end{equation}
and
\begin{equation}
 \phi^{2}(N_{\ast}) = - v^{2}\,W_{-1}\left[-
   \frac{\phi_{end}^{2}}{v^{2}}\left(e^{\frac{N_{\ast}}{\pi}
     +\frac{\phi_{\rm end}^{2}}{m_{\rm Pl}^{2}}}\right)^{-\frac{m_{\rm
         Pl}^{2}}{v^{2}}}\right], 
\label{V0higgs2}
\end{equation}
where $W_{0}$ and $W_{-1}$ correspond to the Lambert functions and
$\phi_{\rm end}$ is given by Eq.~\eqref{phiend_Higgs}. It can be
verified that the solution given by Eq.~(\ref{V0higgs1}) applies when
inflation happens in the small field region of the potential, i.e.,
around the plateau region, $|\phi| < |v|$. The second solution given
by Eq.~(\ref{V0higgs2}), on the other hand, applies in the large field
region of the potential, i.e., when $|\phi| > |v|$.

{}For the Starobinsky potential, Eq.~\eqref{starobinskyV}, the
normalization $V_{0}$ is found to be
\begin{equation}
 \frac{V_{0}^{Staro}}{m_{\rm Pl}^{4}} = \frac{e^{-8\,\sqrt{\frac{
         \pi}{3}}\,\frac{\phi_{\ast}}{m_{\rm Pl}}}}{2 \left(1 -
   e^{-4\sqrt{\frac{\pi}{3}}\frac{\phi_{\ast}}{m_{\rm
         Pl}}}\right)^{4}} \; \Delta_{\rm R},   
\end{equation}
where  $\phi_{\ast}$ is given by
\begin{eqnarray}
\phi(N_*) &=& -\frac{m_{\rm Pl}}{4 \sqrt{3 \pi}} \left\{4N_* +3 +
2\sqrt{3} + \ln \left( -135+ 78 \sqrt{3} \right) \right.  \nonumber
\\ &+& \left. 3 W_{-1}\left[ -\left( 1+ \frac{2}{\sqrt{3}} \right) e^{
    - \frac{4N_*}{3} - 1 - \frac{2}{\sqrt{3}} } \right]  \right\}.
\label{phi*Staro}
\end{eqnarray}


\end{document}